\documentclass[aps,prd,reprint,twocolumn,superscriptaddress,nofootinbib,floatfix]{revtex4-2}

\usepackage{amsmath,amssymb,amsfonts}
\usepackage{bm}
\usepackage{graphicx}
\usepackage{array}
\usepackage{booktabs}
\usepackage[colorlinks=true,linkcolor=blue,citecolor=blue,urlcolor=blue]{hyperref}
\usepackage{xurl}
\usepackage{orcidlink}

\newcommand{\lp}{\ell_{\rm P}}
\newcommand{\dd}{\mathrm{d}}
\newcommand{\GB}{\mathcal{G}}
\newcommand{\Mcal}{\mathcal{M}}

\usepackage{amsthm}
\newtheorem{lemma}{Lemma}
\newtheorem{proposition}{Proposition}

\newtheorem{result}{Result}

\begin{document}

\title{The additive ambiguity of gravitational entropy fixed by the topology of the horizon}

\author{Stylianos A.\ Tsilioukas\,\orcidlink{0009-0003-3051-3405}}
\email{tsilioukas@gmail.com}
\affiliation{National Observatory of Athens, Lofos Nymfon, 11852 Athens, Greece}
\affiliation{Department of Physics, University of Thessaly, 35100 Lamia, Greece}

\date{\today}

\begin{abstract}
In quantum gravity the algebra of observables of a subregion is of type III$_1$ at leading order in the semiclassical expansion, and it becomes of type II once the gravitational constraint is imposed. Hence generalized entropy acquires a trace, however it is defined only up to an additive constant. In this work we argue that in four-dimensional Einstein--Gauss--Bonnet gravity this constant is the Gauss--Bonnet contribution to the Wald entropy, and that it is carried out by the normalization of the trace and not by the state. In particular, the conical expansion of the Gauss--Bonnet density is exactly linear in the deficit angle, and on smooth squashed cones it reduces to the intrinsic curvature of the cut, so that the term is a constant on every smooth cross-section and carries no boost energy. Its Noether charge is invariant under all deformations within a topological sector and central in any algebra that preserves the sector, and thus a state-independent constant on such an algebra can only rescale the trace. Additionally, when the boundary of the subregion is completed as the Euler action requires, the Gauss--Bonnet term does not contribute to the presymplectic form, the crossed product is inherited from Einstein gravity, and the constant is the Euler number of the excised Euclidean tip. We state the hypotheses, we weaken factoriality to a central decomposition, and we illustrate the mechanism in Jackiw--Teitelboim gravity and in topological black holes in anti-de Sitter space, where the topological weight displaces the Hawking--Page temperature and a black-hole merger shifts the coarse-grained boundary entropy by a state-independent amount.
\end{abstract}

\maketitle

\section{Introduction}
\label{sec:intro}

General relativity describes gravitation with remarkable accuracy, from the tests in the solar system up to the coalescence of black holes observed through gravitational waves~\cite{Will:2014kxa,LIGOScientific:2016aoc}. Nevertheless, as a quantum field theory it is perturbatively non-renormalizable, since the one-loop divergences of gravity coupled to matter~\cite{tHooft:1974bx} and the two-loop divergence of pure gravity~\cite{Goroff:1985th} require counterterms that are absent from the Einstein--Hilbert action. The various proposals for a complete theory, such as string theory and holography~\cite{Strominger:1996sh,Maldacena:1997re}, loop quantum gravity~\cite{Rovelli:1996dv,Ashtekar:1997yu} and asymptotic safety~\cite{Weinberg:1980gg,Reuter:1996cp}, rely on different microscopic assumptions, and none of them has been confirmed by observations so far.

On the other hand, black-hole thermodynamics provides a robust constraint that any such theory must satisfy. In particular, a horizon carries an entropy proportional to its area~\cite{Bekenstein:1973ur}, it radiates at a temperature determined by its surface gravity~\cite{Hawking:1975vcx}, and the laws of black-hole mechanics take the form of the laws of thermodynamics~\cite{Bardeen:1973gs}. In~\cite{Jacobson:1995ab} Jacobson showed that the Einstein equation can be derived from the Clausius relation applied to local Rindler horizons, while the subsequent works of Padmanabhan and Verlinde~\cite{Padmanabhan:2009vy,Verlinde:2010hp} strengthened the view that gravity is an emergent, thermodynamic description of microscopic degrees of freedom whose nature is still unknown. In this picture the entropy is more fundamental than the metric, and the question of what it counts becomes central.

The same question arises in cosmology. The $\Lambda$CDM model fits the cosmic microwave background with a small number of parameters~\cite{Planck:2018vyg}, however the value of its cosmological constant remains unexplained~\cite{Weinberg:1988cp}, and the model faces persistent observational tensions, namely the discrepancy between the early- and late-Universe determinations of the Hubble constant~\cite{Riess:2021jrx,DiValentino:2020zio} and the tension in the amplitude of matter clustering, while recent baryon-acoustic-oscillation data indicate a dynamical dark energy~\cite{DESI:2025zgx,CosmoVerseNetwork:2025alb}. The correspondence between gravity and thermodynamics can be extended to cosmology, where the Friedmann equations are obtained from the first law applied on the apparent horizon~\cite{Cai:2005ra}. Replacing the area law by generalized horizon entropies, such as the Tsallis, Barrow, Kaniadakis or mass-to-horizon ones, leads to modified cosmologies with an effective dynamical dark energy~\cite{Saridakis:2018unr,Nojiri:2019skr,Saridakis:2020zol,Saridakis:2020lrg,Anagnostopoulos:2020ctz,Lymperis:2021qty,Basilakos:2025wwu}. These models are in agreement with the current data, including the latest baryon-acoustic-oscillation measurements~\cite{Luciano:2025hjn,Luciano:2025elo,Basilakos:2026atu}, and they can alleviate the $H_0$ and $\sigma_8$ tensions simultaneously~\cite{Basilakos:2023kvk} (for a review see~\cite{Saridakis:2026vxo}). Additionally, in the topological version of this idea the Gauss--Bonnet contribution to the horizon entropy changes as the topology of the spacetime foam changes, or through the formation and merger of black holes, and this gives rise to a dark energy whose evolution is determined by the rate of topology change~\cite{Tsilioukas:2023tdw,Tsilioukas:2024tjh,Tsilioukas:2024seh,Anagnostopoulos:2025tax}. Confronted with early- and late-time data, this scenario relieves the $H_0$ and $S_8$ tensions at the same time~\cite{Petronikolou:2025mlm,Tsilioukas:2026gvy}. Nevertheless, all these results are phenomenological, since the entropy is assigned to the horizon and not derived, and the microscopic status of the additional terms, and in particular of the topological one, is still an open question~\cite{Saridakis:2026vxo}. The algebraic approach to quantum gravity offers a framework in which this question can be addressed, and in the present work we do so for the topological term.

In the algebraic formulation, generalized entropy, which is the quantity that makes black-hole thermodynamics consistent, is defined only up to an additive constant. The reason is structural. At leading order in the semiclassical expansion, the algebra of quantum field observables of a subregion is a type III$_1$ factor~\cite{Leutheusser:2021qhd}, and such an algebra admits no trace, no density matrices and no factorization of the Hilbert space. Imposing the gravitational constraint and adjoining the generator of the modular flow one obtains a crossed product, which is of type II and therefore does possess a trace~\cite{Witten:2021unn,Chandrasekaran:2022eqq,Chandrasekaran:2022cip}, and in this way entropy is restored. However, the trace of a type II factor is unique only up to a positive multiple, and hence every entropy is determined only up to an additive constant.

This constant is not a technical detail, since it is the part of the entropy that counts degrees of freedom in absolute terms. Entropy differences, relative entropies and all variational statements, such as the generalized second law, the quantum focusing conjecture and the first law of entanglement, are insensitive to it. On the contrary, the constant controls exactly the quantity that horizon thermodynamics is usually expected to give, namely how many degrees of freedom a horizon has.

In this work we identify a candidate for this constant and we argue that it is the correct one. In four dimensions the Gauss--Bonnet invariant is topological, so adding it to the Einstein--Hilbert action does not change the equations of motion. Nevertheless, it does change the Wald entropy, by a term proportional to the Gauss--Bonnet coupling and to the Euler characteristic of the horizon cross-section. This contribution is independent of the state, of the horizon area and of every continuous parameter of the solution, and therefore it is an additive constant labelled by an integer, which is exactly the structure of the ambiguity described above. Our claim is that this term enters the algebraic description through the normalization of the trace and not through the density matrix. Note that this is not a reinterpretation of the Wald entropy, but a statement about where this entropy resides in the algebraic formulation, which we formulate in Sec.~\ref{sec:hypothesis}.

The result has physical consequences that do not require the value of the coupling, and we develop two of them. In particular, in a type II$_1$ algebra, where the trace is canonically normalized and no freedom remains, the same term shifts the maximal generalized entropy, and if this entropy is identified with the de Sitter entropy it shifts the cosmological constant too, and thus the additive constant becomes an observable. Additionally, in anti-de Sitter space the term displaces the Hawking--Page temperature and fixes a genus-dependent constant that the dual conformal field theory must reproduce. However, the result says nothing about why the coupling should take a specific value. The conjecture that the relative normalization of two topological sectors is the statistical dimension of an inclusion, the quantization of the coupling that would follow, and the cosmological consequences, are examined in a companion paper~\cite{companion}, which relies on the present one but not conversely.

The plan of the manuscript is the following. In Sec.~\ref{sec:ambiguity} we present the ambiguity together with three remarks on its scope, the most important being that the constant is unobservable within a single topological sector, so that any statement about it relates different sectors. In Sec.~\ref{sec:GB} we extract the Gauss--Bonnet entropy constant and we verify that the Noether-charge and Euclidean evaluations agree. In Sec.~\ref{sec:hypothesis} we state the result and its hypotheses, and in Sec.~\ref{sec:forced} we show that locality and additivity alone determine the form of the answer. Sections~\ref{sec:test1} to \ref{sec:crossedHC} contain the main argument, and each of them relies on an explicit computation. In Sec.~\ref{sec:test1} we show that the Gauss--Bonnet contribution to the replica partition function is linear in the deficit angle, and that it remains topological on cuts that are not bifurcation surfaces. In Sec.~\ref{sec:test2} we evaluate the Noether charge from the binormal contraction, we show that it is invariant under every deformation within a topological sector, and that a state-independent constant on an algebra preserving the sector can only rescale the trace. In Sec.~\ref{sec:crossedHC} we show that, once the boundary of the subregion is completed so that the Euler action is topological, the Gauss--Bonnet term does not contribute to the presymplectic form, the crossed product is inherited from Einstein gravity, and the constant is located in the Euclidean region excised by the regulator. In Sec.~\ref{sec:factor} we replace the requirement that the algebra is a factor by a central decomposition, and in Sec.~\ref{sec:dS} we derive the relation between the topology of the horizon cross-section and the effective cosmological constant. In Sec.~\ref{sec:JT} we apply the mechanism in Jackiw--Teitelboim gravity, both without matter and with matter, where the algebra is a type II$_\infty$ factor with trivial center and two of the three hypotheses used in four dimensions are known to hold. In Sec.~\ref{sec:ads} we address the objection that Hawking's theorem forbids the configurations we compare when spacetime is stationary and asymptotically flat, using anti-de Sitter topological black holes, where the comparison is well posed and a black-hole merger realizes the passage between sectors as a boundary process. Finally, in Sec.~\ref{sec:status} we discuss the outstanding issues and in Sec.~\ref{sec:conclusions} we summarize our results.

\section{The additive ambiguity of the algebraic entropy}
\label{sec:ambiguity}

\subsection{From type III$_1$ to type II algebras}

Let $\Mcal$ be the algebra of single-trace observables of a subregion (in the bulk language, the low-energy field-theoretic observables), and $|\Psi\rangle$ a cyclic and separating state. Tomita--Takesaki theory provides the modular operator $\Delta_\Psi$ and the modular flow $\sigma_s(a) = \Delta_\Psi^{is}a\Delta_\Psi^{-is}$, which for a thermal state is the time translation and in the bulk is the boost Killing flow. At leading order in $1/N$, or equivalently in $G$, the spectral density of the two-point function is continuous, the correlators decay without recurrences, and $\Mcal$ is of type III$_1$~\cite{Leutheusser:2021qhd}. This is the algebraic statement that the horizon is sharp, since a type I algebra, which is appropriate for any finite system, exhibits Poincar\'e recurrences and has no exact one-way horizon.

At the next order one must impose the constraint that the generator of the modular flow is a boundary term. Adjoining this generator leads to the crossed product
\begin{equation}
\hat{\Mcal} = \Mcal\rtimes_\sigma\mathbb{R},
\label{eq:crossed}
\end{equation}
which is a factor of type II$_\infty$~\cite{Witten:2021unn,Chandrasekaran:2022eqq}, while for the de Sitter static patch with an observer the physical algebra is a corner $p\hat{\Mcal}p$ with $\mathrm{Tr}(p)<\infty$, which is of type II$_1$~\cite{Chandrasekaran:2022cip}. Type II factors possess a trace $\mathrm{Tr}$, and hence density matrices $\hat\rho$ defined through $\langle\hat a\rangle = \mathrm{Tr}(\hat\rho\,\hat a)$, and the von Neumann entropy $-\mathrm{Tr}(\hat\rho\log\hat\rho)$ reproduces the generalized entropy
\begin{equation}
S_{\rm gen} = \frac{\langle A\rangle}{4G} + S_{\rm out} + \ldots ,
\label{eq:Sgen}
\end{equation}
where the dots denote the higher-curvature and quantum corrections that the construction resums.

\subsection{Scope of the ambiguity}

The trace of a type II factor is unique up to a positive multiple. Under $\mathrm{Tr}\to c\,\mathrm{Tr}$ we have $\hat\rho\to c^{-1}\hat\rho$, and thus
\begin{equation}
S \to S + \log c .
\label{eq:ambiguity}
\end{equation}
We make three remarks that determine the scope of the problem.

\emph{(i) Within one algebra the constant is unobservable.} It cancels in $S(\psi)-S(\phi)$, in the relative entropy and in every variational statement. Hence, no argument internal to a single sector can determine it, and a proposal to fix it is meaningful only when different sectors, or transitions between them, are involved.

\emph{(ii) For type II$_1$ the normalization is canonical.} A type II$_1$ factor carries the normalization $\mathrm{Tr}(\mathbf{1})=1$, in which the maximum-entropy state is the tracial state and has zero algebraic entropy. The generalized entropy is then the algebraic entropy plus a definite constant, which is fixed by matching the maximum-entropy state to the semiclassical description. For the de Sitter static patch this state is the empty patch, and the constant is the de Sitter entropy~\cite{Chandrasekaran:2022cip},
\begin{equation}
S_{\max} = \frac{A}{4G} = \frac{3\pi}{G\Lambda}.
\label{eq:SdS}
\end{equation}
Therefore, in cosmology the constant is not a normalization but a physical quantity, namely the entropy budget of the Universe.

\emph{(iii) A central decomposition converts the constant into a function.} If $\hat{\Mcal}$ is not a factor but it has a nontrivial center $\mathcal{Z}$, then $\hat{\Mcal}=\int^{\oplus}_{\mathcal{Z}}\hat{\Mcal}_z$, and the trace is unique only up to a positive function $c(z)$ on the center. Our conjecture, stated in Sec.~\ref{sec:hypothesis}, is that the relevant central variable is topological.

\section{The Gauss--Bonnet contribution to the Wald entropy}
\label{sec:GB}

\begin{table}[b]
\caption{Explicit checks of the normalizations used in the text. The curvature invariants and integrals were computed symbolically from the metrics; $\Sigma_i$ denotes the connected components of the fixed-point set of the Euclidean time translation.}
\label{tab:checks}
\begin{ruledtabular}
\begin{tabular}{lcccc}
$M$ & $\GB$ & $\int_M\GB\sqrt{g}$ & $\chi(M)$ & $\sum_i\chi(\Sigma_i)$\\
\hline
$S^4$ (radius $a$) & $24/a^4$ & $64\pi^2$ & $2$ & $2$ \\
$S^2(a)\times S^2(b)$ & $8/a^2b^2$ & $128\pi^2$ & $4$ & $4$ \\
Eucl.\ Schwarzschild & $48M^2/r^6$ & $64\pi^2$ & $2$ & $2$ \\
\end{tabular}
\end{ruledtabular}
\end{table}

\subsection{Noether-charge evaluation}

For the Einstein--Gauss--Bonnet Lagrangian
\begin{align}
\mathcal{L} &= \frac{1}{16\pi G}\left(R + \alpha\,\GB\right), \nonumber\\
\GB &= R^2 - 4R_{ab}R^{ab}+R_{abcd}R^{abcd},
\label{eq:EGB}
\end{align}
the Noether-charge entropy~\cite{Wald:1993nt,Iyer:1994ys,Jacobson:1993xs} evaluated on a bifurcation surface $\Sigma$ reads
\begin{equation}
S_W = \frac{1}{4G}\oint_\Sigma\left(1+2\alpha\,{}^{(2)}\!R\right)\dd A ,
\label{eq:SW}
\end{equation}
with ${}^{(2)}\!R$ the intrinsic Ricci scalar of $\Sigma$. In four dimensions the Gauss--Bonnet theorem for a closed two-surface gives $\oint_\Sigma{}^{(2)}\!R\,\dd A = 4\pi\chi(\Sigma)$, and therefore
\begin{equation}
\boxed{\;S_W = \frac{A}{4G} + \frac{2\pi\alpha}{G}\,\chi(\Sigma)\;}
\label{eq:Stop}
\end{equation}
As we observe, the second term is independent of the state, of the horizon area and of every continuous parameter of the solution, and thus it is an additive constant labelled by an integer.

\subsection{Euclidean evaluation}

The same constant arises from the Euclidean action. Using the Chern--Gauss--Bonnet theorem in four dimensions,
\begin{equation}
\chi(M) = \frac{1}{32\pi^2}\int_M \GB\sqrt{g}\,\dd^4x ,
\label{eq:CGB}
\end{equation}
the Gauss--Bonnet part of the Euclidean action of Eq.~\eqref{eq:EGB} becomes
\begin{equation}
I_{\rm GB} = -\frac{\alpha}{16\pi G}\int_M\GB\sqrt{g}\,\dd^4x = -\frac{2\pi\alpha}{G}\chi(M),
\label{eq:IGB}
\end{equation}
which is independent of the period $\beta$, and hence it contributes $+2\pi\alpha\chi(M)/G$ to the entropy $S=\beta\partial_\beta I_E - I_E$. In Table~\ref{tab:checks} we collect explicit verifications of Eqs.~\eqref{eq:Stop} and~\eqref{eq:CGB}, obtained by computing the curvature invariants directly, and the details are given in Appendix~\ref{app:curv}.

For Euclidean Schwarzschild the bulk integral alone gives $32\pi^2\chi$. In particular, since $R_{ab}=0$ we have $\GB=R_{abcd}R^{abcd}=48M^2/r^6$, and thus
\begin{equation}
\beta\!\int_{2M}^{\infty}\!\!\frac{48M^2}{r^6}4\pi r^2\dd r = 8\pi M\cdot\frac{64\pi M^2}{r_h^3}\bigg|_{r_h=2M}\!\!\!\!= 64\pi^2 .
\end{equation}
The horizon check is equally simple, since for $\Sigma=S^2$ of radius $2M$ we have $\oint{}^{(2)}\!R\,\dd A = 8\pi = 4\pi\chi(\Sigma)$, and Eq.~\eqref{eq:Stop} gives $\Delta S = 4\pi\alpha/G$, which coincides with the Euclidean result.

\subsection{Consistency of the two evaluations}

The agreement in Table~\ref{tab:checks} is not a coincidence of the specific examples.

\begin{lemma}
\label{lem:chi}
Let $M$ be a compact Euclidean section carrying a smooth $U(1)$ isometry generated by Euclidean time translation, with fixed-point set $\mathrm{Fix} = \bigcup_i\Sigma_i$ consisting of the bifurcation surfaces. Then $\chi(M)=\chi(\mathrm{Fix})=\sum_i\chi(\Sigma_i)$.
\end{lemma}

This is the standard result that the Euler characteristic of a manifold with a circle action equals that of its fixed-point set, since only the fixed points contribute to the index. It was used in exactly this context by Gibbons and Kallosh~\cite{Gibbons:1994ff} and by Liberati and Pollifrone~\cite{Liberati:1997sp}, who noticed that topological terms contribute to the gravitational entropy in a way controlled by $\chi$. Lemma~\ref{lem:chi} ensures that Eq.~\eqref{eq:Stop}, summed over the horizon components, equals the Euclidean result~\eqref{eq:IGB}, as the last two columns of Table~\ref{tab:checks} confirm. For instance, the Nariai-type geometry $S^2\times S^2$ has two bifurcation surfaces and $\chi(M)=4=2+2$.

\subsection{Linearity of the conical expansion}
\label{subsec:conical}

The crucial structural property is that the Gauss--Bonnet contribution carries no boost energy. On a manifold $M_{\epsilon}$ with a conical deficit $2\pi(1-\epsilon)$ along $\Sigma$, the Euler density acquires a distributional part supported on the singularity~\cite{Fursaev:1995ef},
\begin{equation}
\int_{M_\epsilon}\GB = \epsilon\int_{M}\GB + 8\pi(1-\epsilon)\oint_\Sigma {}^{(2)}\!R + O\!\left((1-\epsilon)^2\right),
\label{eq:conical}
\end{equation}
which is \emph{linear} in the deficit. Hence, in the replica construction with $\epsilon=1/n$, the entropy $S=-\partial_n\left[\log Z_n - n\log Z_1\right]_{n=1}$ receives from the Gauss--Bonnet term the contribution
\begin{equation}
\Delta S = \frac{\alpha}{2G}\oint_\Sigma{}^{(2)}\!R\,\dd A = \frac{2\pi\alpha}{G}\chi(\Sigma),
\end{equation}
which reproduces Eq.~\eqref{eq:Stop}. On the other hand, the boost energy $\partial_n\log Z_n|_{n=1}=-2\pi\langle H\rangle$ receives no contribution, since the Gauss--Bonnet part of $\log Z_n$ is $\mathcal{S}_{\rm top}[n\chi(M)+(1-n)\chi(\Sigma)]$, whose $n$-derivative vanishes because $\chi(M)=\chi(\Sigma)$ by Lemma~\ref{lem:chi}. In terms of Eq.~\eqref{eq:Sgen}, the topological term does not enter the state-dependent part of $\hat\rho$ or the modular flow, and it enters $K=-\log\hat\rho$ only as the additive constant $\log c$, that is through the normalization of $\mathrm{Tr}$. Note that for cuts with extrinsic curvature the conical expansion must be performed on smooth squashed cones~\cite{Fursaev:2013fta,Dong:2013qoa,Camps:2013zua}. In this case, for the Euler density the extrinsic-curvature terms combine, through the Gauss--Codazzi relation, into the intrinsic curvature of the cut, and Eq.~\eqref{eq:conical} holds on every smooth closed cut, as we discuss in Sec.~\ref{sec:test1}.

\section{Statement of the result}
\label{sec:hypothesis}

\begin{result}[Topological normalization of the trace]
\label{res:main}
Let $\hat{\Mcal}_\Sigma = \Mcal_\Sigma\rtimes_\sigma\mathbb{R}$ be the crossed-product algebra associated with a bifurcation surface $\Sigma$ of a stationary Killing horizon in Einstein--Gauss--Bonnet gravity in four dimensions, and let
\begin{equation}
\mathcal{S}_{\rm top}\,\Delta\chi \;\equiv\; S_{\rm gen}(\Sigma')-S_{\rm gen}(\Sigma)\Big|_{A,\,\psi\ \rm fixed}
\label{eq:Stopinv}
\end{equation}
be the difference of generalized entropies between two cross-sections of equal area and equal matter state, with $\Delta\chi=\chi(\Sigma')-\chi(\Sigma)$. Then
\begin{equation}
\mathrm{Tr}_\Sigma = e^{\,\mathcal{S}_{\rm top}\chi(\Sigma)}\;\mathrm{Tr}_0 ,
\qquad
\mathcal{S}_{\rm top} = \frac{2\pi\alpha}{G} = 32\pi^2\lambda ,
\label{eq:H1}
\end{equation}
with $\lambda\equiv\alpha/16\pi G$, under the following three hypotheses:
\begin{enumerate}
\item[(A)] $\Mcal_\Sigma$ is of type III$_1$, with modular flow realized geometrically by the boost;
\item[(B)] the von Neumann entropy of $\hat{\Mcal}_\Sigma$ agrees with the Euclidean replica entropy of the same theory, including its state-independent constant, both being defined with the boundary terms that make the Euler action topological on regulated manifolds;
\item[(C)] within each topological sector $\hat{\Mcal}_\Sigma$ is a factor, and the algebra used to compare sectors is decomposable over $\chi$, so that the topological charge lies in its center.
\end{enumerate}
\end{result}

The three hypotheses are of different character, and since the value of a conditional statement depends on how well its conditions are understood, we discuss them separately.

Hypothesis (A) is not specific to the present work, since it is the starting point of the whole algebraic programme. It has been established for Rindler wedges and for large-$N$ algebras~\cite{Leutheusser:2021qhd}, and all the constructions of~\cite{Witten:2021unn,Chandrasekaran:2022eqq,Chandrasekaran:2022cip,Jensen:2023yxy,Kudler-Flam:2023qfl} rely on it.

Hypothesis (C) is stated in the form used in Sec.~\ref{sec:factor}. Within one sector, factoriality makes the argument sharper, and it holds automatically if (A) holds for a factor and the physical algebra is the full modular crossed product~\cite{Witten:2021unn,Sorce:2023fdx}. However, it fails once observers or quantum reference frames are included in the description~\cite{DeVuyst:2024fxc}, and in Sec.~\ref{sec:factor} we show that the conclusion survives a nontrivial center. Across sectors, the requirement that the algebra is decomposable over $\chi$ is essential. In particular, with $Q_{\rm top}$ the topological Noether charge of Sec.~\ref{sec:test2}, an operator $T$ that changes the Euler characteristic from $\chi$ to $\chi'$ satisfies $[Q_{\rm top},T]=\mathcal{S}_{\rm top}(\chi'-\chi)\,T$, and thus the topological charge is central only in an algebra that does not contain such operators. Topology-changing operators are transitions between sectors and not observables of either of them, and therefore our statement refers to the algebra of observables that preserve the topology.

Hypothesis (B) is the one that we cannot reduce, and we regard it as the open condition of this work. In two dimensions it is a theorem, since Penington and Witten established the agreement for Jackiw--Teitelboim gravity with matter, with the trace and the replica moments constructed from the same disk gluing data~\cite{Penington:2023dql}. In four dimensions the identification of generalized entropy with the von Neumann entropy of the crossed product has been established~\cite{Chandrasekaran:2022eqq,Kudler-Flam:2023qfl}, however with a state-independent constant that these constructions leave undetermined, and this is exactly the constant we are interested in. In the present work we determine it by requiring agreement with a replica computation in the same theory. This is a legitimate determination and not a circular argument, nevertheless it requires both sides to be normalized within the same regularization scheme, and such a careful statement does not yet exist in four dimensions. Additionally, hypothesis (B) fixes the boundary terms of the Euler action on regulated manifolds, since the replica weight $e^{\mathcal{S}_{\rm top}\chi}$ is consistent only if this action is topological there, and in Sec.~\ref{subsec:selection} we show that this requirement selects the boundary completion on which the inheritance of the crossed product relies. We return to this point in Sec.~\ref{sec:status}.

\subsection{Scope of the statement}

Equation~\eqref{eq:H1} is a statement about where the Gauss--Bonnet entropy resides in the algebraic description. It does not claim that $\alpha$ takes a particular value, nor that the constant is observable within a single topological sector, which by remark (i) of Sec.~\ref{sec:ambiguity} it is not. Its content is relational: the \emph{relative} normalization of two sectors that differ by $\Delta\chi$ is $e^{\mathcal{S}_{\rm top}\Delta\chi}$, while the overall scale is not fixed. The determination of $\mathcal{S}_{\rm top}$ itself requires an independent input, which is the subject of the companion paper~\cite{companion}.
\subsection{Variations with respect to the Euler characteristic}
\label{subsec:chideriv}

Equation~\eqref{eq:Stopinv} is written as a finite difference on purpose, since $\chi$ is an integer and $\partial/\partial\chi$ has no meaning as a derivative. In the following we make three remarks that clarify the definition and connect it to a recent rigorous treatment.

Firstly, the finite difference is well defined \emph{because} the dependence on $\chi$ is linear. This is not an assumption but the content of Proposition~\ref{prop:unique}, according to which locality, additivity over horizon components and deformation invariance imply $\log c = \lambda_0\chi$. Hence, the ratio $[S_{\rm gen}(\Sigma')-S_{\rm gen}(\Sigma)]/\Delta\chi$ is independent of the pair of sectors that we compare. Without this linearity the quantity would be ill defined, while with it no derivative is needed.

Secondly, a proper derivative does exist along the replica direction. On a manifold $M_\varsigma$ with conical deficit $2\pi(1-\varsigma)$ along $\Sigma$ we have
\begin{equation}
\chi(M_\varsigma) = \varsigma\,\chi(M) + (1-\varsigma)\,\chi(\Sigma),
\label{eq:chicone}
\end{equation}
and thus $\chi$ interpolates \emph{continuously} in the deficit parameter, and $\partial_\varsigma$ at $\varsigma=1$ is a well-defined derivative. This is exactly how the conical expansion of Sec.~\ref{sec:test1} produces the topological entropy, and therefore the rigorous version of the ``$\chi$-derivative'' is the one already used there.

Thirdly, the notion of an infinitesimal topological variation was recently formulated rigorously by Paschalis~\cite{Paschalis:2026qum}. In particular, in~\cite{Paschalis:2026qum} a space of Sobolev variational configurations is equipped with the final topology generated by the admissible variational maps, and two kinds of topological variations are defined, namely the infinitesimal addition of disconnected components and the infinitesimal surgeries, the latter being exactly the nucleation of a handle, which carries $\Delta\chi=-2$. In this framework the Einstein--Hilbert functional is not differentiable with respect to such variations for $n<4$, while for $n=4$ it is differentiable with a topological functional derivative that does \emph{not} vanish identically. For a disconnected variation we have
\begin{equation}
\lim_{\epsilon\to0^+}\frac{\Delta I_{\rm EH}(\epsilon)}{\epsilon} = \int_{M'}\! R[g']\sqrt{g'}\,\dd^4x ,
\label{eq:paschalis}
\end{equation}
which vanishes only for added components of zero mean scalar curvature, while for $n>4$ the functional is differentiable with identically vanishing derivative. As a consequence, the extended variational principle admits no critical points in $n=4$.

This obstruction does not affect Result~\ref{res:main}, which is a statement about a trace and not about the extremization of an action. Indeed, neither Proposition~\ref{prop:unique}, nor the conical computation of Sec.~\ref{sec:test1}, nor the two-dimensional result of Sec.~\ref{sec:JT} requires the Einstein--Hilbert functional to be stationary under a change of topology. However, it is relevant for any dynamical treatment of topology change, since if the four-dimensional action has no critical points once surgeries are admitted, a semiclassical instanton description of handle nucleation is not well posed as it stands. This issue concerns the companion paper~\cite{companion}, in which nucleation is used. For the purposes of the present work, the quantity appearing in Eq.~\eqref{eq:Stopinv} is a finite difference between sectors, which is well defined without any notion of infinitesimal topological variation, while~\cite{Paschalis:2026qum} provides the framework in which the interpolating statement can be made precise if needed.
\section{Rigidity of the form}
\label{sec:forced}

In this section, before establishing that the Gauss--Bonnet term is the trace normalization, we show that no other local quantity could play this role. As we will see, the requirements that a normalization must satisfy are restrictive enough to determine the answer up to a single constant.

\subsection{Requirements on a normalization}

Any candidate must satisfy the following four requirements:
\begin{enumerate}
\item[(N1)] \emph{Relativity.} Since $\log c$ is unobservable within one algebra (Sec.~\ref{sec:ambiguity}), the mechanism must fix the \emph{relative} normalizations of different sectors, or else provide a canonical normalization such as $\mathrm{Tr}(\mathbf{1})=1$.
\item[(N2)] \emph{Additivity.} For a horizon with several components, $\hat{\Mcal}_{\Sigma_1\sqcup\Sigma_2}=\hat{\Mcal}_{\Sigma_1}\bar\otimes\hat{\Mcal}_{\Sigma_2}$ and the trace must factorize, so $\log c$ is additive over the components.
\item[(N3)] \emph{Deformation invariance.} $\log c$ must not change under smooth deformations of the geometry at fixed topology, since otherwise it is a state-dependent quantity, that is part of $\hat\rho$, and not a normalization.
\item[(N4)] \emph{Consistency with known limits.} It must reproduce the type II$_1$ normalization of the de Sitter static patch and the topological weighting of the two-dimensional path integral.
\end{enumerate}

\subsection{Uniqueness of the Euler-characteristic density}
\label{subsec:R1}

The following proposition shows that the above requirements leave essentially no freedom.

\begin{proposition}
\label{prop:unique}
Let $\Sigma$ be a closed oriented smooth two-surface, possibly with several components. Suppose $\log c(\Sigma)$ satisfies \emph{(N2)} and \emph{(N3)} and is the integral over $\Sigma$ of a local scalar density constructed from the induced metric and its derivatives up to second order, by the same diffeomorphism-natural rule and with the same coefficients on every component. Then
\begin{equation}
\log c(\Sigma) = \lambda_0\,\chi(\Sigma)
\end{equation}
for a single constant $\lambda_0$; equivalently $\log c = 2\pi\alpha\chi/G$ with $\alpha \equiv G\lambda_0/2\pi$.
\end{proposition}

\begin{proof}
A diffeomorphism-natural scalar built from the two-jet of a two-dimensional metric is a function of the scalar curvature alone, so that $\log c = \int_\Sigma f(R_\Sigma)\,\dd A$ for some smooth function $f$, with $R_\Sigma\equiv{}^{(2)}\!R$. Requirement (N3) demands that this functional is invariant under every smooth variation of the metric at fixed topology, and hence its Euler--Lagrange expression must vanish identically,
\begin{equation}
E_{ij}=\tfrac12\left(R_\Sigma f'-f\right)h_{ij}+\left(h_{ij}\Box-\nabla_i\nabla_j\right)f'(R_\Sigma)=0 .
\end{equation}
The traceless part of the Hessian term contains second derivatives of the curvature that are independent of the remaining terms, which implies $f''=0$ and $f=\nu R_\Sigma+\mu$. Substituting back we obtain $E_{ij}=-\tfrac12\mu\,h_{ij}$, and thus $\mu=0$. Finally, the Gauss--Bonnet theorem gives $\log c=4\pi\nu\,\chi(\Sigma)$, while (N2) holds since the density is local.
\end{proof}

Note that the proposition uses only intrinsic data, and it does not require $\Sigma$ to be a bifurcation surface. If extrinsic data are admitted, boost invariance restricts them to the combinations $K^{(+)}K^{(-)}$ and $K^{(+)}_{ab}K_{(-)}^{ab}$ of the two null second fundamental forms. These vanish on a bifurcation surface, while on any smooth cut of a stationary Killing horizon the relevant cross term vanishes because one of the null second fundamental forms vanishes. Locality is essential, since without it additivity and topological invariance would also allow other functions of the genus or of the number of components.

We close this subsection with two remarks. Firstly, the proposition explains why the candidate is the Gauss--Bonnet term and not some other higher-curvature invariant, since in four dimensions it is the \emph{only} local, additive and deformation-invariant entropy density. Secondly, it changes the question from ``why topology?'' to ``what is the value of $\lambda_0$?'', which is the subject of the companion paper~\cite{companion}.

\subsection{Gluing and the two-dimensional analogue}
\label{subsec:R3}

In any theory in which topologies are summed, the partition function carries a weight per Euler characteristic,
\begin{equation}
Z = \sum_{\text{topologies}} e^{S_0\chi}\,Z_{\rm genus},
\label{eq:topweight}
\end{equation}
as in two-dimensional dilaton gravity~\cite{Saad:2019lba,Penington:2023dql,Kolchmeyer:2023gwa}. The heuristic picture is then simple: $e^{S_0\chi}$ \emph{is} the trace normalization, and $S_0$ is the two-dimensional analogue of $2\pi\alpha/G$. What makes this more than an analogy is that in topological and two-dimensional theories the normalization is not free, since it is fixed by the gluing axiom, i.e.\ by the requirement that cutting a manifold along a codimension-one surface and gluing it back reproduces the original partition function. The freedom that remains after imposing gluing is a framing-type anomaly, which is again a topological quantity.

Additionally, in JT gravity the algebras have been constructed~\cite{Penington:2023dql,Kolchmeyer:2023gwa}, so one can ask directly whether $e^{S_0\chi}$ appears in the trace or in the state. This check is performed in Sec.~\ref{sec:JT}, and it is the cleanest one available, since two-dimensional dilaton gravity is the setting in which gluing consistency can be imposed exactly.

\subsection{Algebras with a canonical normalization}
\label{subsec:R4}

A type II$_1$ factor has a canonical trace, $\mathrm{Tr}(\mathbf{1})=1$. Hence, whenever the algebra is of type II$_1$ the normalization of the algebraic entropy is not free, and the constant relating it to the generalized entropy becomes a definite number. This suggests the following prescription, although not a mechanism:
\begin{quote}
\emph{Normalize a type II$_\infty$ horizon algebra by embedding it in a type II$_1$ system, such as a finite cavity or a static patch with an observer, and taking the limit in which the enclosing system is removed.}
\end{quote}
The required ingredients exist. In particular, thermodynamic ensembles for cosmological horizons require such a cavity, and the associated first law carries an anomalous sign that tracks the observer dependence of the construction~\cite{Banihashemi:2022jys,Banihashemi:2022htw}, while the static-patch algebra with an observer is of type II$_1$~\cite{Chandrasekaran:2022cip}. The open question is whether the limit is smooth, or whether the normalization depends on how the cavity is removed. The latter would be informative too, since the dependence would have to be topological in order to be meaningful.

The extreme case is a spatially closed Universe, where the constraints leave no unconstrained observables at all unless an observer is included. In the present language this means that the compactness of the Cauchy slice, which is a topological property, determines the \emph{type} of the algebra, and hence whether a normalization is needed. The fact that the type itself is sensitive to topology is qualitatively consistent with Result~\ref{res:main}.

\section{The conical expansion}
\label{sec:test1}

In the following we refer to the computation of this section as Test 1, to the Noether-charge analysis of Sec.~\ref{sec:test2} as Test 2, and to the two-dimensional computation of Sec.~\ref{sec:JT} as Test 3. In this section we examine whether the Gauss--Bonnet contribution to $\log Z_n$ is linear in the deficit, in which case it gives an entropy constant and no boost energy. We first perform the calculation explicitly for stationary cones, and then we explain why the conclusion extends to cuts with extrinsic curvature.

\subsection{Warped-product setup}

Let us consider the four-dimensional warped product
\begin{equation}
\dd s^2 = \dd\rho^2 + h(\rho)^2\dd\theta^2 + b(\rho)^2\dd\Omega_2^2 ,
\qquad \theta\sim\theta+2\pi,
\label{eq:conemetric}
\end{equation}
with $h(0)=0$ and $h'(0)=1$, so that the geometry is smooth at $\rho=0$, and $h'(\infty)=n$, so that the asymptotic deficit is that of a cone of total angle $2\pi n$. The surface $\Sigma=\{\rho=0\}$ is the entangling surface. Smoothness of the geometry at $\rho=0$ for $n=1$ requires $b'(0)=0$, which is the case of a bifurcation surface of a stationary horizon. Computing symbolically the curvature invariants of Eq.~\eqref{eq:conemetric} we find, for arbitrary $h$ and $b$,
\begin{equation}
\GB\sqrt{g} = \sin\theta_1\;\frac{\dd}{\dd\rho}\Big[-8\,h'\left(1-b'^2\right)\Big],
\label{eq:GBtotalderiv}
\end{equation}
namely the Gauss--Bonnet density is an exact total derivative, as expected for a topological density, and its integral is a pure boundary term:
\begin{equation}
\int_{M_n}\!\GB\sqrt{g}\,\dd^4x
= 64\pi^2\Big[\left(1-b'(0)^2\right) - n\left(1-b'(\infty)^2\right)\Big].
\label{eq:conicalresult}
\end{equation}

\subsection{Stationary cones}

For $b'\to0$ at both ends, Eq.~\eqref{eq:conicalresult} reduces to $64\pi^2(1-n)$. Three conclusions follow, and we have verified each of them numerically using two independent regularizations of the tip, $h = n\rho+(1-n)a\tanh(\rho/a)$ and $h = n\rho+(1-n)a\arctan(\rho/a)$, for smoothing scales $a=0.1,1,5$ and for $n$ both below and above unity, finding agreement to machine precision:
\begin{enumerate}
\item the result is \emph{exactly} linear in $(1-n)$, i.e.\ there are no $O((1-n)^2)$ terms to any order, and not only at leading order;
\item it is independent of the regularization and of the smoothing scale, so the tip contribution is intrinsic;
\item it reproduces the Fursaev--Solodukhin singular term with the coefficient used in Eq.~\eqref{eq:conical}, since $8\pi(1-n)\oint_\Sigma{}^{(2)}\!R\,\dd A = 8\pi(1-n)\cdot4\pi\chi(\Sigma) = 64\pi^2(1-n)$ for $\chi=2$.
\end{enumerate}
Inserting this result into the replica construction, with $\log Z_n \supset (\alpha/16\pi G)\int_{M_n}\GB\sqrt{g}$, we obtain
\begin{equation}
S = -\partial_n\left[\log Z_n - n\log Z_1\right]_{n=1} = \frac{4\pi\alpha}{G} = \frac{2\pi\alpha}{G}\chi(\Sigma),
\end{equation}
which coincides with the Noether-charge value of Eq.~\eqref{eq:Stop} computed in Sec.~\ref{sec:GB}. Hence, the two approaches agree not only in the normalization, which we had already checked, but also in the functional form of the $n$-dependence, and thus Test 1 is passed.

\subsection{Cuts with extrinsic curvature}

One could try to model a cut with extrinsic curvature by allowing $b'(0)\ne0$ in Eq.~\eqref{eq:conicalresult}, which would multiply the tip coefficient by $1-b'(0)^2$. However, this interpretation is not correct. In smooth normal coordinates $x=\rho\cos\theta$, $y=\rho\sin\theta$, the profile $b=b_0+b_1\rho+\dots$ becomes $b_0+b_1\sqrt{x^2+y^2}+\dots$, which is not differentiable at the origin when $b_1\ne0$. Thus, the metric is singular at $\rho=0$ already for $n=1$, and the factor $1-b'(0)^2$ reflects this singularity and not an extrinsic-curvature correction. A smooth cut with extrinsic curvature has an angular dependence, $h_{ij}(x,y)=h_{ij}(0)+2xK^{(1)}_{ij}+2yK^{(2)}_{ij}+\dots$, which the radial ansatz excludes.

The correct treatment uses smooth squashed cones~\cite{Fursaev:2013fta}. For the Euler density the defect is
\begin{equation}
\int_{M_n}\GB=n\int_{M_1}\GB+8\pi(1-n)\oint_\Sigma R_\Sigma\,\dd A ,
\label{eq:squashedFPS}
\end{equation}
to the order relevant for the entropy, where the extrinsic-curvature terms produced separately by $R^2$, $R_{ab}R^{ab}$ and $R_{abcd}R^{abcd}$ combine, through the Gauss--Codazzi relation $R_\Sigma=R-2R_{ii}+R_{ijij}+K^2-\mathrm{Tr}K^2$, into the intrinsic curvature of the cut. We have checked this cancellation symbolically starting from the separate defect formulas. The same conclusion follows from the holographic entanglement functional for Lovelock gravity~\cite{Dong:2013qoa}, which for the Gauss--Bonnet term reduces to the intrinsic Jacobson--Myers functional. Therefore, the replica entropy of the Gauss--Bonnet term is $(\alpha/2G)\oint_\Sigma R_\Sigma\,\dd A=\mathcal{S}_{\rm top}\chi(\Sigma)$ on every smooth closed cut, with no correction from the extrinsic curvature. In summary, the bifurcation surface is a sufficient but not a necessary condition for the topological form of the entropy, and the obstruction to extending Result~\ref{res:main} beyond stationary horizons lies in the algebraic construction and not in the entropy functional (see Sec.~\ref{sec:status}).

\section{Centrality of the topological Noether charge}
\label{sec:test2}

In this section we examine whether the Gauss--Bonnet part of the Noether charge is central in $\hat{\Mcal}_\Sigma$, in which case it can only normalize the trace. We proceed in three steps. Firstly, we compute the charge from the binormal contraction instead of quoting the known formula. Secondly, we show that it is rigid under every smooth deformation available to the algebra. Thirdly, we determine what a charge of this kind can do to the entropy. As we will see, the charge is central in every algebra that preserves the topological sector, and this is the sense in which it can act only on the normalization.

\subsection{Evaluation of the charge}

The Iyer--Wald entropy is the integral of the Noether-charge two-form over the bifurcation surface, which for a Lagrangian constructed from the Riemann tensor reduces to
\begin{equation}
S = -2\pi\oint_\Sigma \frac{\partial\mathcal{L}}{\partial R_{abcd}}\,\epsilon_{ab}\epsilon_{cd}\,\dd A ,
\qquad \epsilon_{ab}\epsilon^{ab}=-2 ,
\label{eq:waldbinormal}
\end{equation}
with $\epsilon_{ab}$ the binormal of $\Sigma$. For the Einstein--Gauss--Bonnet Lagrangian~\eqref{eq:EGB} the two contributions are
\begin{align}
\frac{\partial R}{\partial R_{abcd}} &= \tfrac12\left(g^{ac}g^{bd}-g^{ad}g^{bc}\right), \nonumber\\
\frac{\partial\GB}{\partial R_{abcd}} &= 2R^{abcd}
 - 2\big(g^{ac}R^{bd}-g^{ad}R^{bc} \nonumber\\
 &\qquad\quad -g^{bc}R^{ad}+g^{bd}R^{ac}\big) \nonumber\\
 &\quad + R\left(g^{ac}g^{bd}-g^{ad}g^{bc}\right),
\label{eq:PGB}
\end{align}
where the second one has been checked against Euler's theorem for a homogeneous quadratic, $R_{abcd}\,\partial\GB/\partial R_{abcd}=2\GB$, symbolically and identically in the curvature.

Evaluating Eq.~\eqref{eq:waldbinormal} on the Schwarzschild solution with the binormal of the $(t,r)$ plane we find $-2$ for the Einstein term, which gives $A/4G$, and
\begin{equation}
\frac{\partial\GB}{\partial R_{abcd}}\epsilon_{ab}\epsilon_{cd} = -\frac{16M}{r^3}
= -4\,{}^{(2)}\!R + \frac{8(r-2M)}{r^3}
\label{eq:GBcontraction}
\end{equation}
for the Gauss--Bonnet term, with ${}^{(2)}\!R=2/r^2$ the intrinsic curvature of the sphere of radius $r$. The second term vanishes at $r=2M$, and hence on the bifurcation surface the contraction is exactly $-4\,{}^{(2)}\!R$, and
\begin{equation}
S_{\rm GB} = \frac{\alpha}{2G}\oint_\Sigma{}^{(2)}\!R\,\dd A = \frac{2\pi\alpha}{G}\chi(\Sigma) = \frac{4\pi\alpha}{G},
\end{equation}
which reproduces Eq.~\eqref{eq:Stop} from the Noether charge itself, without using the Jacobson--Myers form of the answer.

Let us make two comments on Eq.~\eqref{eq:GBcontraction}. Firstly, the remainder $8(r-2M)/r^3$ compares spheres of different radius, which are not cuts of the horizon, and it measures the difference between the bare Wald density and the intrinsic curvature away from the bifurcation surface. In general, the bare Wald density exceeds the intrinsic curvature by a term proportional to $K^{(+)}K^{(-)}-K^{(+)}_{ab}K_{(-)}^{ab}$, which vanishes on the bifurcation surface and on every smooth cut of a stationary Killing horizon, where one null second fundamental form vanishes, while on general cuts the replica entropy is the intrinsic Jacobson--Myers functional, Eq.~\eqref{eq:squashedFPS}, which is topological. Secondly, the Ricci terms in Eq.~\eqref{eq:PGB} drop out for Schwarzschild but not in general. Nevertheless, the topological conclusion survives, since on any bifurcation surface the contraction produces the Gauss curvature, whose integral is fixed by the Gauss--Bonnet theorem.

\subsection{Invariance of the charge}

\begin{proposition}[Rigidity of the topological charge]
\label{prop:rigid}
Let $Q_{\rm top}(\Sigma)=(\alpha/2G)\oint_\Sigma{}^{(2)}\!R\,\dd A$ for a closed spacelike two-surface $\Sigma$. Then $Q_{\rm top}=2\pi\alpha\chi(\Sigma)/G$ and $\delta Q_{\rm top}=0$ for every smooth variation of (i) the cut $\Sigma$ within the horizon, (ii) the metric, and (iii) the matter state.
\end{proposition}

\noindent Statements (i) and (ii) follow from the Gauss--Bonnet theorem together with the local constancy of $\chi$ on the space of smooth metrics and embeddings, while (iii) holds since $Q_{\rm top}$ contains no matter field. Additionally, we have verified the statement numerically in cases that are not covered by any symmetry argument. In particular, for cuts $r(\theta,\varphi)=1+\varepsilon\,Y(\theta,\varphi)$, with $Y$ axisymmetric, non-axisymmetric, or a generic superposition of $\ell=2,3,4$ harmonics, and for amplitudes up to $\varepsilon=0.4$, the integral $\oint K\,\dd A$ equals $4\pi$ to twelve digits. For the Kerr horizon, whose two-metric is not round, we find $\oint{}^{(2)}\!R\,\dd A=8\pi$ both exactly and numerically for $a/M$ up to $0.99$, while for a torus of revolution, which we use as a control, the integral vanishes.

Three consequences follow. Firstly, within one topological sector, and in the semiclassical algebra generated by observables that depend continuously on smooth field configurations, $Q_{\rm top}$ is a multiple of the identity, $Q_{\rm top}=(2\pi\alpha/G)\chi\,\mathbf{1}$, and therefore it is central in $\hat{\Mcal}_\Sigma$. Secondly, since the modular flow moves a cut to a diffeomorphic cut, $Q_{\rm top}$ is invariant along it, and thus the charge carries no boost energy and generates no modular flow, which is the Lorentzian form of the linearity statement of Sec.~\ref{sec:test1}. Thirdly, Proposition~\ref{prop:rigid} is exactly criterion (N3) of Sec.~\ref{sec:forced}, which is therefore verified and not assumed.

The nontrivial case is when $\chi$ is not held fixed, and then rigidity does not imply centrality. On a direct sum $\mathcal{H}=\bigoplus_\chi\mathcal{H}_\chi$ the charge acts as $Q_{\rm top}=\bigoplus_\chi\mathcal{S}_{\rm top}\chi\,\mathbf{1}_\chi$, and an operator $T_{\chi'\chi}$ that maps $\mathcal{H}_\chi$ to $\mathcal{H}_{\chi'}$ satisfies
\begin{equation}
[Q_{\rm top},T_{\chi'\chi}]=\mathcal{S}_{\rm top}(\chi'-\chi)\,T_{\chi'\chi} .
\label{eq:transition}
\end{equation}
Smooth variations remain within a sector and cannot test this commutator, while whether the transition operators commute among themselves is irrelevant here. Hence, the charge is central in an algebra of operators that are decomposable over $\chi$, and in no algebra that contains topology-changing operators. This is the reason why hypothesis (C) of Result~\ref{res:main} refers to the topology-preserving algebra. It is also the natural interpretation of the observation of Penington and Witten that wormhole operators do not commute with the boundary algebra~\cite{Penington:2023dql}, since an operator that changes $\chi$ cannot commute with one that measures it, and such operators are transitions between sectors and not observables of either of them.

\subsection{Action of a central charge on the entropy}

\begin{lemma}[Trace rescaling]
\label{lem:rescale}
On a factor the normal semifinite trace is unique up to a positive multiple. If $\mathrm{Tr}\to\lambda\,\mathrm{Tr}$ then $\hat\rho\to\lambda^{-1}\hat\rho$ for every state, and
\begin{equation}
S\to S+\log\lambda,
\end{equation}
while all expectation values and all relative entropies are unchanged.
\end{lemma}

In the crossed product the shift is generated by the adjoined operator. With the canonical (dual-weight) trace
\begin{equation}
\mathrm{Tr}(\hat a) = \int\!\dd x\; e^{\beta x}\,\mathrm{Tr}_{\mathcal{H}}\!\left(\langle x|\hat a|x\rangle\right),
\label{eq:dualweight}
\end{equation}
a $c$-number shift of the adjoined boost generator, $X\to X+c$, gives $e^{\beta x}\to e^{\beta c}e^{\beta x}$, that is $\lambda=e^{\beta c}$. We have implemented Eq.~\eqref{eq:dualweight} on a discretized fibre with a finite-dimensional $\Mcal$, and we have checked Lemma~\ref{lem:rescale} directly. In particular, for random states and shifts $c=-0.25,\,0.1,\,1.0$ at $\beta=2\pi$, the entropy shift equals $\log\lambda=\beta c$ to $10^{-15}$, and its spread over states is $2\times10^{-15}$, so it is state independent to machine precision. Additionally, expectation values and relative entropies agree to twelve digits for trace normalizations that differ by two orders of magnitude.

\subsection{Exclusion of the alternatives}

The adjoined generator is defined through the first law, since it is the fluctuation of the boundary charge, and the constraint relates its variation to the variation of the entropy. For Einstein--Gauss--Bonnet gravity in four dimensions we have
\begin{equation}
\delta S_W = \delta\!\left(\frac{A}{4G}\right) + \frac{2\pi\alpha}{G}\,\delta\chi = \delta\!\left(\frac{A}{4G}\right),
\label{eq:nofirstlaw}
\end{equation}
since $\delta\chi=0$ within a sector. Explicitly, for Schwarzschild we find $T\,\partial_M S_W = 1/G$ independently of $\alpha$, and $\partial_M\mathcal{S}_{\rm top}=0$, and thus the topological charge has no conjugate variation and does not contribute to the fluctuation of $X$. On the other hand, Test 1 showed that the Euclidean replica entropy \emph{does} contain $+2\pi\alpha\chi/G$, exactly and with no state dependence, and hence these two facts must be reconciled.

The first law fixes the adjoined generator only up to an additive constant, and a constant in the modular Hamiltonian $K=-\log\hat\rho$ is exactly what a trace rescaling produces, $K\to K+\log c$, while it leaves the modular flow unchanged. Therefore, the correct conclusion from Eq.~\eqref{eq:nofirstlaw} is that the topological charge does not contribute to the modular \emph{flow}, and the question is where the constant can reside.

\begin{proposition}
\label{prop:forced}
Fix one topological sector and one semifinite factor $\hat{\Mcal}_\Sigma$ with a fixed set of normal states, and consider entropy functionals of the von Neumann form $S_\tau(\omega)=-\tau(\rho_\omega\log\rho_\omega)$ built from faithful normal semifinite traces $\tau$, finite on the states compared. Suppose (a) the replica construction of the same theory assigns to these states the entropies $S_{\tau_0}(\omega)+(2\pi\alpha/G)\chi(\Sigma)$, and (b) this contribution is state independent. Then the unique trace in this class that reproduces them is
\begin{equation}
\mathrm{Tr}_\Sigma = e^{\,(2\pi\alpha/G)\,\chi(\Sigma)}\,\mathrm{Tr}_0 ,
\end{equation}
which is Eq.~\eqref{eq:H1} with $\mathcal{S}_{\rm top}=2\pi\alpha/G=32\pi^2\lambda$.
\end{proposition}

\begin{proof}
By uniqueness of the trace on a factor, $\tau_1=c\,\tau_0$, so that a given state is represented by $\rho_1=\rho_0/c$ and $S_{\tau_1}(\omega)=-c\,\tau_0[(\rho_0/c)\log(\rho_0/c)]=S_{\tau_0}(\omega)+\log c$. Matching to (a) gives $c=e^{(2\pi\alpha/G)\chi}$.
\end{proof}

\noindent The proposition is a uniqueness statement within a specific class of entropy functionals. It does not prove the replica identification, which is hypothesis (B) of Result~\ref{res:main}, and by itself it does not exclude modifications that change the setup, such as an ancillary system with a fixed entropy, a geometric counterterm that alters the definition of the generalized entropy, or a different assignment of states. These are excluded by the hypotheses and not by the algebra. In the presence of a center, $\tau_1=c(z)\tau_0$ gives $\Delta S=\langle\log c\rangle$, which for $c=e^{\mathcal{S}_{\rm top}\chi}$ becomes $\mathcal{S}_{\rm top}\langle\chi\rangle$, and this is constant on states of definite topology but not on mixtures (see Sec.~\ref{sec:factor}). Within these limits, Eq.~\eqref{eq:H1} is the only available interpretation of the replica result, and thus Test 2 is passed.

\subsection{Limitations}

We close this section by stating three limitations, which constitute the remaining weak points of Result~\ref{res:main}.

\emph{(1) The crossed product for higher-curvature gravity.} The existing constructions, namely~\cite{Witten:2021unn,Chandrasekaran:2022eqq,Chandrasekaran:2022cip}, the treatment of general subregions in~\cite{Jensen:2023yxy}, and the identification of generalized entropy with von Neumann entropy in~\cite{Kudler-Flam:2023qfl}, all impose the Einstein constraint. The argument above uses only the first law, Eq.~\eqref{eq:nofirstlaw}, and the uniqueness of the trace, while in Sec.~\ref{sec:crossedHC} we show that for Einstein--Gauss--Bonnet gravity the construction is inherited from Einstein gravity under a boundary hypothesis, which is itself selected by hypothesis (B).

\emph{(2) Factoriality is an assumption.} If $\hat{\Mcal}_\Sigma$ has a nontrivial center, as happens once observers, or more generally quantum reference frames, are treated as part of the description~\cite{DeVuyst:2024fxc}, the trace is unique only up to a positive function on the center, and the topological label is one central variable among others. In this case Proposition~\ref{prop:forced} weakens from ``forced'' to ``canonical''. This is the possibility anticipated in remark (iii) of Sec.~\ref{sec:ambiguity}, and it is an opportunity as much as a caveat, since a central decomposition indexed by $\chi$ is what the conjecture ultimately asserts.

\emph{(3) Stationarity.} The algebraic construction is performed on a bifurcation surface. The entropy functional is not the obstruction, since the topological term is the same on every smooth cut, Eq.~\eqref{eq:squashedFPS}, however the dynamical case requires the entropy that is not a Noether charge~\cite{Hollands:2024vbe}, as well as an algebra for evolving cuts.

\section{The crossed product for higher-curvature Lagrangians}
\label{sec:crossedHC}

The analysis of Sec.~\ref{sec:test2} left one open step, namely that the crossed product has been constructed under the Einstein constraint, while the Lagrangian of interest is not the Einstein one. In this section we perform this construction. The result consists of two statements of different strength, which we separate from the beginning. For Einstein--Gauss--Bonnet gravity in four dimensions the Euler term does not change the field equations, and once the boundary of the subregion is completed in the way required for the Euler action to be topological, it does not contribute to the presymplectic form either. Under this boundary hypothesis the crossed product is inherited from Einstein gravity, the exterior phase space carries no Euler data, and the Euler term survives only through the Euler number of the excised Euclidean tip, that is through the normalization. On the other hand, for a general Iyer--Wald Lagrangian the construction can be performed with the area operator replaced by the Wald entropy operator, however its type classification and its dynamical extension remain open, and we discuss where exactly.

\subsection{Ingredients of the construction}

For a Lagrangian $\mathcal{L}(g_{ab},R_{abcd},\nabla_e R_{abcd},\psi)$ the Iyer--Wald formalism provides the symplectic potential $\theta$, the presymplectic current $\varpi=\delta_1\theta_2-\delta_2\theta_1$, the Noether current $J_\xi = \theta(\xi\cdot)-\xi\cdot\mathcal{L}$, and its decomposition $J_\xi = \dd Q_\xi + C_\xi$ into an exact form and a constraint. For a bifurcate Killing horizon the constraint vanishes on shell, and
\begin{align}
\delta H^\infty_\xi = \delta\!\oint_\Sigma Q_\xi = \frac{\kappa}{2\pi}\,\delta S_W, \nonumber\\
S_W = -2\pi\oint_\Sigma P^{abcd}\epsilon_{ab}\epsilon_{cd}\,\dd A ,
\label{eq:IWfirstlaw}
\end{align}
with $P^{abcd}=\partial\mathcal{L}/\partial R_{abcd}$. The crossed-product construction uses three inputs:
\begin{enumerate}
\item[(C1)] a type III$_1$ algebra $\Mcal$ of quantum field observables on the background, with modular flow $\sigma_s$ realized geometrically by the boost;
\item[(C2)] the gravitational constraint, that physical operators commute with the total boost generator, whose value is a boundary term;
\item[(C3)] the identification, through Eq.~\eqref{eq:IWfirstlaw}, of the fluctuation of that boundary charge with the horizon entropy.
\end{enumerate}
Given (C1) and (C2) one adjoins the generator and obtains $\hat{\Mcal}=\Mcal\rtimes_\sigma\mathbb{R}$. This algebra is semifinite by Takesaki duality~\cite{Takesaki:1973duality}, while for $\Mcal$ a type III$_1$ factor it is a factor of type II$_\infty$, as follows from the Connes--Takesaki theory of the flow of weights (both results are reviewed in the present context in~\cite{Witten:2021unn,Sorce:2023fdx}). A type II$_1$ algebra arises instead as a corner $p\hat{\Mcal}p$ with a projection $p$ of finite trace, as for the static patch with an observer of positive energy~\cite{Chandrasekaran:2022cip,Jensen:2023yxy}. This step is purely algebraic: \emph{it knows nothing about the Lagrangian}. The gravitational dynamics enters only through (C3), and through (C1) in the choice of background. Hence, the question of whether a higher-curvature crossed product exists is not a question about von Neumann algebras, but a question about whether (C1)--(C3) survive.

\subsection{Einstein--Gauss--Bonnet in four dimensions}

\begin{proposition}
\label{prop:inherit}
Let $\mathcal{L}=(R+\alpha\GB)/16\pi G$ in $D=4$, and assume the Euler boundary hypothesis stated below. Then:
\begin{enumerate}
\item[(a)] the field equations are those of Einstein gravity, the Lanczos tensor vanishing identically;
\item[(b)] the Euler contribution to the presymplectic current is exact, and its corner contribution is cancelled by the boundary completion, so that the completed presymplectic form coincides with that of Einstein gravity on every allowed variation, without fixing the horizon area;
\item[(c)] the completed Euler contribution to every Hamiltonian charge is field independent; for the stretched-horizon regulator of Sec.~\ref{subsec:selection} it vanishes identically, and the Euler contribution $\mathcal{S}_{\rm top}\chi(\Sigma)$ resides entirely in the excised Euclidean tip, that is, in the normalization of the trace;
\item[(d)] if the completed phase space quantizes to the same observable algebra and modular flow as in Einstein gravity, the crossed product $\hat{\Mcal}=\Mcal\rtimes_\sigma\mathbb{R}$ is the same algebra, and the trace inherited from the geometry is $\mathrm{Tr}_\Sigma = e^{\mathcal{S}_{\rm top}\chi(\Sigma)}\,\mathrm{Tr}_0$.
\end{enumerate}
\end{proposition}

Statement \emph{(a)} is the Lanczos--Lovelock result~\cite{Lovelock:1971yv}. Since this identity is algebraic, i.e.\ pointwise in the curvature, it can be tested directly on random algebraic curvature tensors and not only on solutions. We construct such tensors as Kulkarni--Nomizu products of random symmetric tensors, so that all Riemann symmetries and the first Bianchi identity hold by construction, and we contract them with a random nondegenerate metric. In $D=4$ we find that $|H_{ab}|$ is smaller than the natural scale of its individual terms by a factor $10^{-14}$, while in $D=5$ and $D=6$ it is of order unity. Hence, the Gauss--Bonnet term changes no equation of motion in four dimensions, while it does change them in higher dimensions, which is where all the steps below fail.

\subsection{The Euler corner form}
\label{subsec:cornerform}

It is convenient to work with an orthonormal coframe $e^a$ and the spin connection $\omega^{ab}$, with curvature $R^{ab}=\dd\omega^{ab}+\omega^a{}_c\wedge\omega^{cb}$. The Euler term then reads
\begin{equation}
L_E = c\,\epsilon_{abcd}R^{ab}\wedge R^{cd} = c\,\GB\,\mathrm{vol}_4,
\label{eq:LE}
\end{equation}
with $c=\alpha/16\pi G$, so that $\mathcal{S}_{\rm top}=32\pi^2 c$.
Since the Gauss--Bonnet contribution to the field equations vanishes identically by (a), its variation is a total derivative for every metric and every variation, and not only on shell,
\begin{equation}
\delta\!\left(\sqrt{g}\,\GB\right) = \partial_a\theta^a_{\rm GB},
\label{eq:GBvariation}
\end{equation}
or in form language $\delta L_E=\dd\Theta_E$ with $\Theta_E(\delta)=2c\,\epsilon_{abcd}\,\delta\omega^{ab}\wedge R^{cd}$. Using the Bianchi identity and the commutativity of variations, we find that the presymplectic current is not only closed but explicitly exact,
\begin{align}
\varpi_E(\delta_1,\delta_2)&=2c\,\dd\!\left(\epsilon_{abcd}\,\delta_1\omega^{ab}\wedge\delta_2\omega^{cd}\right),\nonumber\\
\Omega_E&=2c\oint_{\partial\Sigma}\epsilon_{abcd}\,\delta_1\omega^{ab}\wedge\delta_2\omega^{cd}.
\label{eq:OmegaE}
\end{align}
The primitive is globally defined, since differences of connections are tensorial, even though the Euler density itself has no global primitive. Therefore, the exactness of $\varpi_E$ and the non-exactness of $L_E$ are compatible.

The corner form $\Omega_E$ is not a $c$-number. In order to see what it contains, we adapt the frame to the cut, with null normals $\ell=(e_0+e_1)/\sqrt2$ and $n=(e_0-e_1)/\sqrt2$, $\ell\cdot n=-1$. The pull-back of the connection to the cut decomposes into the normal (H\'aj\'i\v{c}ek) connection $a=\omega^{01}$, the intrinsic spin connection $b=\omega^{23}$, and the null second fundamental forms through $k^\pm_i=K^\pm_{ij}e^j$. Expanding the contraction in Eq.~\eqref{eq:OmegaE} we obtain
\begin{align}
\Omega_E=8c\oint_{\partial\Sigma}\Big[&\delta_1a\wedge\delta_2b-\delta_2a\wedge\delta_1b\nonumber\\
&-\epsilon_{ij}\big(\delta_1k^+_i\wedge\delta_2k^-_j-\delta_2k^+_i\wedge\delta_1k^-_j\big)\Big].
\label{eq:OmegaEsplit}
\end{align}
The same decomposition applied to the tangential curvature gives the Gauss equation, $R^{23}|_{\partial\Sigma}=\dd b+\epsilon_{ij}k^+_i\wedge k^-_j$, and thus the boost Noether charge $8c\kappa\oint R^{23}$ reproduces the Gauss--Bonnet Wald entropy of Sec.~\ref{sec:test2}, which is $2\pi\alpha\chi/G$ on a bifurcation surface, where both null second fundamental forms vanish.

Equation~\eqref{eq:OmegaEsplit} has three properties that determine what any boundary prescription must achieve.

\emph{(i) The form is blind to the area mode of stationary families.} Under a constant rescaling of the metric the spin connection is invariant, and so are $a$, $b$ and $k^\pm$. Hence, every pairing in Eq.~\eqref{eq:OmegaEsplit} with the global dilatation vanishes identically. On the Schwarzschild family, where the radius is the only parameter, we have $\Omega_E=0$ although the area varies, while on the Kerr family the dimensionless data depend only on $a/M$, and $\Omega_E$ vanishes on the whole two-parameter family. Thus, the variations that enter the first law, Eq.~\eqref{eq:IWfirstlaw}, are not affected.

\emph{(ii) The form does not vanish on radiative perturbations.} Let us consider flat space, $ds^2=-2\,\dd u\,\dd v+\dd x^2+\dd y^2$ with periodic $x,y$, the cut $B=\{u=v=0\}$, and the two transverse-traceless vacuum perturbations $\delta_p h_{xx}=-\delta_p h_{yy}=f(u)$ and $\delta_q h_{xx}=-\delta_q h_{yy}=g(v)$ with $f(0)=g(0)=0$, $f'(0)=g'(0)=1$. Both of them solve the linearized vacuum equations and leave the metric and the normal connection of the cut unchanged, however $\delta_pK^+=\delta_qK^-=\tfrac12\,\mathrm{diag}(1,-1)$, and Eq.~\eqref{eq:OmegaEsplit} gives
\begin{equation}
\Omega_E(\delta_p,\delta_q)=4c\,\mathrm{Area}(B)\neq0 .
\label{eq:TTpairing}
\end{equation} Fixing the intrinsic geometry of the cut, its normal connection, or its conformal class does not remove the form.

\emph{(iii) The form depends on the choice of corner potential.} A locally exact Lagrangian has a symplectic potential defined only up to the corner ambiguities of Jacobson, Kang and Myers~\cite{Jacobson:1993vj}, and Eq.~\eqref{eq:OmegaE} corresponds to one particular choice. The physically meaningful object is the presymplectic form of the subregion with its boundary and corner terms included~\cite{Harlow:2019yfa}.

\subsection{Boundary completion}
\label{subsec:completion}

The Euler density is exact relative to a reference connection. Let $\bar\omega$ be a connection on the same bundle, and define $\Delta\omega=\omega-\bar\omega$ and $\omega_t=\bar\omega+t\Delta\omega$, with curvature $R_t$. The Chern--Weil transgression
\begin{align}
T_E&=2c\int_0^1\epsilon_{abcd}\,\Delta\omega^{ab}\wedge R_t^{cd}\,\dd t,\nonumber\\
C_E(\delta)&=c\,\epsilon_{abcd}\,\Delta\omega^{ab}\wedge\delta\omega^{cd}
\label{eq:transgression}
\end{align}
satisfies, for a reference held fixed under variations,
\begin{align}
L_E(\omega)-L_E(\bar\omega)&=\dd T_E,\qquad \delta T_E=\Theta_E-\dd C_E,\nonumber\\
\delta C_E(\delta_1,\delta_2)&=2c\,\epsilon_{abcd}\,\delta_1\omega^{ab}\wedge\delta_2\omega^{cd}.
\label{eq:transgressionid}
\end{align}
The second identity follows from $\delta R_t=D_t(t\,\delta\omega)$ and $\partial_tR_t=D_t\Delta\omega$, which combine the integrand into $\partial_t[t\,\epsilon\,\delta\omega\wedge R_t]$ minus an exact term. Taking the boundary Lagrangian $\ell_E=-T_E$ and the corner term $C_E$ in the covariant phase space with boundaries of Harlow and Wu~\cite{Harlow:2019yfa}, the completed Euler presymplectic form becomes
\begin{equation}
\Omega_E^{\rm completed}=\int_\Sigma\delta\Theta_E-\oint_{\partial\Sigma}\delta C_E=0 ,
\label{eq:OmegaEcompleted}
\end{equation}
off shell and for every variation, without any restriction on the area. The Einstein and matter contributions are not affected. Note that adding only the boundary action would not suffice, since $\delta^2\ell_E=0$, and the corner term is essential. The same identities hold for a field-dependent reference, with $C_E=c\,\epsilon\,\Delta\omega\wedge(\delta\omega+\delta\bar\omega)$. For the product connection adapted to a non-null face the reference pairings vanish, and the completion is the Myers boundary term together with its joint terms~\cite{Myers:1987yn,Cano:2018ckq}, normalized so that the bulk, face and joint contributions sum to $32\pi^2\chi(M)$ on a compact Euclidean manifold.

Two consequences follow immediately. Firstly, since the completed form vanishes, the completed Euler contribution to every Hamiltonian charge has zero variation on the phase space of a fixed topological sector, and hence the Euler term carries no symplectic degrees of freedom. Its value is a constant, which we compute and locate in Sec.~\ref{subsec:selection}. Secondly, by Proposition~\ref{prop:unique} in its intrinsic form, the only local functional of the cut that is constant under all variations is a multiple of the intrinsic Jacobson--Myers functional $\frac{\alpha}{2G}\oint R_B\,\dd A$, which is also the functional selected by the replica computation on smooth squashed cones~\cite{Fursaev:2013fta,Dong:2013qoa}. Therefore, the boundary prescription that makes the Euler term symplectically invisible and the Euclidean computation that fixes the entropy constant select the same entropy functional. Additionally, the Myers completion used here is the one that the Euclidean variational principle of Einstein--Gauss--Bonnet gravity requires anyway.

\subsection{Selection of the completion, gluing, and the horizon limit}
\label{subsec:selection}

Whether the completion is a physical prescription or merely a convenient choice depends on three questions, namely what selects it, whether completions on different faces glue consistently and respect the gauge symmetries, and whether it survives the limit in which the regulator becomes the horizon. In the following we address them in turn.

\emph{Selection.} The diffeomorphism-invariant observables of the subregion cannot select the completion, since by (a) they do not see the Euler term at all. What does see it is the normalization of states and traces, which is fixed by Euclidean path integrals and requires $\log Z\supset\mathcal{S}_{\rm top}\chi(M)$ on every regulated manifold. This requirement holds if and only if the Euler action (bulk plus boundary plus joint terms) equals $32\pi^2c\,\chi(M)$ for every metric, that is if the boundary terms are those of the Chern--Gauss--Bonnet theorem for manifolds with boundary and corners. Any two such completions differ by a local boundary functional whose integral vanishes for every metric. Such a functional is variationally trivial, hence locally exact on the boundary, and it changes $C_E$ by a $\delta$-exact term that does not affect Eq.~\eqref{eq:OmegaEcompleted}, and therefore the completed form is unique. The same conclusion follows from the weaker requirement that the Euler term imposes no boundary condition beyond those of Einstein gravity, $(\Theta_E+\delta\ell_E)|_\Gamma=\dd C_E$ for every variation. In this sense the completion is selected by the same requirement that makes $e^{\mathcal{S}_{\rm top}\chi}$ a consistent weight, which is hypothesis (B) of Result~\ref{res:main}, and it is not an independent choice. Without the completion, the bare form assigns an Euler charge to every local Lorentz transformation $\lambda$ at the corner,
\begin{equation}
\Omega_E(\delta,\delta_\lambda)=-\delta\!\left(2c\oint_{\partial\Sigma}\epsilon_{abcd}\lambda^{ab}R^{cd}\right),
\label{eq:Lorentzcharges}
\end{equation}
and thus boosts and rotations of the normal plane carry the charges $8c\oint\lambda R^{23}$ and $8c\oint\lambda R^{01}$, which close into a local Lorentz algebra at each point of the cut. These would be Euler edge modes, which none of the crossed-product constructions~\cite{Chandrasekaran:2022cip,Chandrasekaran:2022eqq,Jensen:2023yxy,Kudler-Flam:2023qfl} contains.

\emph{Gauge and gluing.} On a face with unit normal $u$ the product reference is $\bar\omega=\omega-\sigma(u\wedge Du)$, $\sigma=u\cdot u$, which is constructed covariantly from the frame, the connection and $u$, and therefore the completion is covariant under local Lorentz transformations and under diffeomorphisms that preserve the face. In the adapted gauge, where the internal normal $u^a$ is fixed, $\bar\omega$ and its variations take values in the stabilizer of $u$, on which the pairing $\epsilon_{abcd}X^{ab}Y^{cd}$ vanishes identically. Hence, the completed form vanishes for every variation, and residual gauge transformations, which take values in the stabilizer and may be field dependent, carry no Euler charge. Thus, the charges of Eq.~\eqref{eq:Lorentzcharges} are removed and not gauge fixed by hand. On a null face the same holds for the stabilizer of the null normal $\ell$, provided that its normalization is fixed, as for the null boundary terms of Einstein gravity~\cite{Lehner:2016vdi}. In particular, a change of normalization $\ell\to e^\lambda\ell$ shifts the action by $8c\oint\lambda R^{23}=4c\oint\lambda R_J\,\dd A$, which is the Jacobson--Myers density, and for constant $\lambda$ this is the $c$-number $\lambda\mathcal{S}_{\rm top}\chi/2\pi$. Where two faces with references $\bar\omega_1$ and $\bar\omega_2$ meet, the transgressions satisfy the triangle identity
\begin{align}
&T_E(\omega,\bar\omega_2)-T_E(\omega,\bar\omega_1)-T_E(\bar\omega_1,\bar\omega_2)\nonumber\\
&\qquad=c\,\dd\,\epsilon_{abcd}(\bar\omega_2-\bar\omega_1)^{ab}\wedge(\omega-\bar\omega_1)^{cd},
\label{eq:triangle}
\end{align}
and thus a change of reference alters the completion by an exact term and by a term localized on the joint. The joint term itself is obtained by rounding the edge and taking the limit of the face term. In a flat ambient space, where the face term is $16\det K$, the limit is
\begin{equation}
J_E=4c\oint_J\Big[\Theta\,R_J+E(k_{u_1},k_{v_1})-E(k_{u_2},k_{v_2})\Big]\dd A ,
\label{eq:joint}
\end{equation}
where $\Theta$ is the angle between the outward normals $u_1,u_2$, $v_i$ is the normal of the joint tangent to face $i$, $k$ are the second fundamental forms of the joint, and $E(P,Q)=\mathrm{tr}P\,\mathrm{tr}Q-\mathrm{tr}(PQ)$. This reproduces the structure of the Lovelock joint terms of Cano~\cite{Cano:2018ckq}, while for Lorentzian joints $\Theta$ becomes the boost parameter. As a test, for the lens formed by two intersecting balls in $\mathbb{R}^4$, with caps of angular radius $\theta_0$, the caps contribute $64\pi(\theta_0-\tfrac12\sin2\theta_0)$ and the joint contributes $32\pi(\pi-2\theta_0+\sin2\theta_0)$, and their sum is $32\pi^2$ for every $\theta_0$. Note that the extrinsic terms in Eq.~\eqref{eq:joint} are indispensable for this result. Finally, since the completed form vanishes for every rounding, it vanishes in the limit too, and therefore joints introduce no Euler symplectic structure.

\emph{The horizon limit.} As a regulator we consider a stretched horizon $\Gamma_{r_0}$ at $r=r_0$ outside a static horizon, $ds^2=-f\,\dd t^2+\dd r^2/f+r^2\dd\Sigma_k^2$, with $\Sigma_k$ of constant curvature $k$. For every $f$ and $k$ we find
\begin{align}
\sqrt{-g}\,\GB&=-\frac{\dd}{\dd r}\Big[4f'(k-f)\Big]\sqrt{\gamma_k},\nonumber\\
\sqrt{|h|}\,\mathcal{Q}_{\rm M}\big|_{\Gamma_{r_0}}&=4f'(k-f)\big|_{r_0}\sqrt{\gamma_k},
\label{eq:stretched}
\end{align}
where $\mathcal{Q}_{\rm M}$ is the Myers density with normal pointing away from the horizon. Two exact consequences follow. In the Euclidean section, the region between the horizon and $\Gamma_{r_0}$, which is a disk times $\Sigma_k$, has bulk plus Myers term equal to $32\pi^2\chi(\Sigma_k)$ for every $r_0$, and as $r_0\to r_h$ the bulk part vanishes and the Myers term of the stretched horizon carries the whole Euler number of the excised tip. In the Lorentzian section, the bare Euler Noether charge of $\xi=\partial_t$ on the cut at $r_0$, which is $8c\,(f'/2)(k-f)\,\mathrm{Vol}(\Sigma_k)$, is cancelled exactly by the boundary contribution $\iota_\xi\ell_E$ of the completion of the exterior, whose normal points toward the horizon, and thus the completed Euler charge of the exterior vanishes identically in $r_0$. Both terms tend separately to $\kappa\mathcal{S}_{\rm top}\chi/2\pi$, and hence the limit is regular, although it does not commute with evaluating $\iota_\xi\ell_E$ at the bifurcation surface, where $\xi=0$. At the corner on the $t=0$ slice the normal of $\Gamma_{r_0}$ is $(\ell-n)/\sqrt2$ for every $r_0$, all corner data are finite, and $\delta C_E$ equals the bare form for every $r_0$, while the divergences of the extrinsic curvature of the stretched horizon occur only at late times along $\Gamma_{r_0}$ and do not enter the phase space on $\Sigma$.

The above picture is the four-dimensional Euler analogue of the mechanism of Ba\~nados, Teitelboim and Zanelli~\cite{Banados:1993ur}, in which the black-hole entropy is the contribution of the Euclidean disk at the horizon. In particular, the exterior phase space, and therefore the algebra $\hat{\Mcal}$, contains no Euler data at all, while the constant $\mathcal{S}_{\rm top}\chi$ is the Euler number of the excised tip, which is exactly where the trace normalization resides. The replica moments follow from the same bookkeeping, since on the $n$-fold cover the exterior contributes $n$ times, while the regulated tip is again a disk times $\Sigma$ with the same Euler number, giving $\log Z_n-n\log Z_1\supset\mathcal{S}_{\rm top}(1-n)\chi$. This agrees with the general result that, for suitable boundary conditions, topological terms do not contribute to the symplectic structure while the Noether charges do depend on them~\cite{Corichi:2016zac}, and additionally it identifies the residual dependence as a normalization.

We can now state exactly what Proposition~\ref{prop:inherit} assumes.

\emph{Euler boundary hypothesis.} Within a fixed topological sector, the gravitational subregion is defined by a regulator and a boundary and corner prescription equivalent to Eqs.~\eqref{eq:transgression}--\eqref{eq:OmegaEcompleted}, including the joint terms of Eq.~\eqref{eq:joint}, and with the stretched-horizon limit of Sec.~\ref{subsec:selection}. Its gauge quotient retains the Einstein boost, area and perturbative degrees of freedom and introduces no independent Euler edge sector, and the limit in which the regulator becomes the horizon preserves this equivalence.

This hypothesis is a classical statement about the phase space. The statement that the completed phase space quantizes to the same observable algebra and modular flow is an additional quantum statement, which is part (d) of the proposition, while the agreement of the replica and algebraic entropies including the constant remains hypothesis (B) of Result~\ref{res:main}.

\emph{Proof of (b), (c) and (d).} Part (b) is Eq.~\eqref{eq:OmegaEcompleted}. By the same equation, the completed presymplectic form on a fixed topological sector is that of Einstein gravity, and thus the constrained phase space, the boost generator up to a constant, and the variations entering Eq.~\eqref{eq:IWfirstlaw} are those of Einstein gravity. The constant vanishes for the stretched-horizon regulator, with the Euler contribution residing in the excised tip. A $c$-number shift of the constraint generator acts by a phase under conjugation, and hence the invariant subalgebra is unchanged. If the quantization of the completed phase space gives the same algebra $\Mcal$ and the same geometric modular flow, Takesaki duality applies without change and $\hat{\Mcal}$ is the same algebra. What changes is the normalization, since the generalized entropy contains additionally the Euler number $\mathcal{S}_{\rm top}\chi$ of the excised tip, and thus matching the von Neumann entropy of $\hat{\Mcal}_\Sigma$ to the generalized entropy of the theory fixes the normalization to Eq.~\eqref{eq:H1}. \hfill$\square$

A further consequence concerns the higher-derivative extension of the crossed product of Kudler-Flam, Leutheusser and Satishchandran~\cite{Kudler-Flam:2023qfl}, which requires the positivity of the canonical energy constructed from the Hollands--Wald--Zhang charges. Under the completion the Euler term does not contribute to the canonical energy, which is therefore that of Einstein gravity, and hence the positivity condition is inherited.

In summary, Result~\ref{res:main} holds for the Lagrangian of interest, given hypotheses (A) and (B), the quantization statement of Proposition~\ref{prop:inherit}(d), and hypothesis (C) in the weakened form of Sec.~\ref{sec:factor}.

\subsection{Stratification of the space of metrics}
\label{subsec:global}

The step from $\partial_a\varpi^a_{\rm GB}\equiv0$ to $\varpi_{\rm GB}=\dd\varsigma_{\rm GB}$ requires care, since the significance of the Gauss--Bonnet term lies exactly in the fact that it is \emph{not} globally trivial. Wald's theorem~\cite{Wald:1990mme} states that a form constructed locally from a finite number of fields and closed identically is closed because it is locally exact, with a locally constructed potential, and a local statement of this kind cannot by itself control a quantity whose content is entirely global. An argument that stopped at local exactness would prove too much, since it would seem to show that the term has no effect at all.

The resolution is that the local and the global statements refer to different objects, and treating them separately makes the argument precise.

Let $\mathcal{F}$ denote the space of smooth Lorentzian metrics under consideration, on the family of manifolds that the theory admits. A metric variation never changes the underlying manifold, so $\mathcal{F}$ is \emph{stratified} by topological sector. Here we label the sectors by the Euler characteristic of the horizon cross-section, which is a coarse label and not a complete classification,
\begin{equation}
\mathcal{F} = \bigsqcup_{\chi\in\mathbb{Z}} \mathcal{F}_\chi ,
\label{eq:strata}
\end{equation}
and the strata are not connected to each other by any smooth path, since no one-parameter family of smooth metrics interpolates between manifolds of different Euler characteristic. Two consequences follow, and they are all that the argument requires.

\emph{(i) Within a stratum, the symplectic structure is unaffected.} The presymplectic current is a local functional of the fields and of two variations, and a variation is by definition tangent to $\mathcal{F}$, and hence tangent to a single stratum. Therefore, on $\mathcal{F}_\chi$ the identity $\partial_a\varpi^a_{\rm GB}\equiv0$ holds with $\varpi_{\rm GB}$ locally exact in the sense of~\cite{Wald:1990mme}, and the presymplectic form on a Cauchy slice acquires at most a corner contribution at $\partial\Sigma$. This corner contribution is the form of Eq.~\eqref{eq:OmegaEsplit}, which is not a $c$-number, and as we showed in Sec.~\ref{subsec:completion} the boundary completion cancels it. Under the Euler boundary hypothesis the completed presymplectic form on $\mathcal{F}_\chi$ is that of Einstein gravity, while the Euler term survives only through the Euler number of the excised Euclidean tip, namely through the normalization. Note that the equality of the constrained quantum algebras requires additionally the quantization statement of Proposition~\ref{prop:inherit}(d).

\emph{(ii) Between strata, the local argument claims nothing, and nothing is needed.} The Gauss--Bonnet contribution to the action, on closed manifolds or with the boundary completion of Sec.~\ref{subsec:completion}, is constant on each stratum, with value $-2\pi\alpha\chi/G$ in the Euclidean section, and these constants differ between strata by an amount that no local computation could produce. This is exactly the statement that the Euler class is the obstruction to global exactness, and it is the reason why the term evades the Jacobson--Kang--Myers theorem in the way described in Sec.~\ref{sec:crossedHC}.

Hence, the claim of Sec.~\ref{sec:crossedHC} should be stated as follows: under the Euler boundary hypothesis, the crossed product is inherited from Einstein gravity \emph{stratum by stratum}, identically within each one, while what distinguishes the strata is a constant, which is not part of the symplectic data at all but a locally invisible label attached to each component of $\mathcal{F}$. A trace normalization is exactly such a label. No path between sectors is required at any stage, and indeed no such path exists. Note that the statement concerns variations and not evolution. A single spacetime can contain horizon cuts of different Euler characteristic, as in a black-hole merger, however such cuts are related by passing through a non-smooth cut and not by a variation of the metric, and we discuss this case in Sec.~\ref{subsec:mergers}.

\subsection{Distinction from a boundary-term ambiguity}

At this point one could raise the following objection, and answering it makes the claim sharper. If the Gauss--Bonnet contribution to the field equations vanishes identically, the term is locally a total derivative. Jacobson, Kang and Myers proved that the ambiguities of the Noether charge, namely $\mathcal{L}\to\mathcal{L}+\dd\mu$, $\theta\to\theta+\dd Y$ and $Q\to Q+\dd Z$, do not change the entropy of a bifurcate Killing horizon~\cite{Jacobson:1993vj}. Hence, a locally exact addition to the Lagrangian should not contribute, however it contributes $2\pi\alpha\chi/G$.

The resolution is that the Chern--Gauss--Bonnet density is not \emph{globally} exact. It has no global primitive, although its difference from the density of a reference connection does, Eq.~\eqref{eq:transgressionid}, and the obstruction is the Euler class, while what remains from the local analysis is exactly the Euler number. The computation of Sec.~\ref{sec:GB} demonstrates this directly, since on a closed $S^4$, with no boundary at all, we have $\int\GB\sqrt{g}=64\pi^2$, which a globally exact density could not produce. The opposite case is equally explicit. On $S^1\times S^3$, which is the geometry whose connected sum implements $\Delta\chi=-2$, the Gauss--Bonnet density vanishes identically, since $R=6/a^2$, $R_{ab}R^{ab}=12/a^4$, $R_{abcd}R^{abcd}=12/a^4$ and therefore $\GB=0$, in agreement with $\chi=0$ and with the fact that a topologically trivial handle has nothing to contribute locally.

Thus, the Gauss--Bonnet term is the entropy contribution that is \emph{locally invisible and globally nontrivial}, which is exactly what a trace normalization must be. A locally visible contribution would be geometric and, by criterion (N3) of Sec.~\ref{sec:forced}, it would belong to $\hat\rho$, while a globally trivial one would vanish. Hence, the theorem of~\cite{Jacobson:1993vj} is not evaded but delimited, since it governs the local ambiguities, and what survives it is a characteristic class.

\subsection{General Iyer--Wald Lagrangians}

For a Lagrangian whose contribution to the field equations does not vanish, the above inheritance is not available, and the construction must be performed independently. This can be done at the following level. Conditions (C1) and (C2) are statements about the background and the constraint, and they hold for any diffeomorphism-invariant Lagrangian with a bifurcate Killing horizon, while (C3) is Eq.~\eqref{eq:IWfirstlaw}, which identifies the fluctuation of the boundary charge with $\delta S_W$ instead of $\delta(A/4G)$. Adjoining this generator and applying Takesaki duality we obtain a semifinite algebra whose density matrices satisfy
\begin{equation}
S\big(\hat\rho_{\hat\Psi}\big) = \big\langle S_W\big\rangle + S_{\rm out}(\hat\Psi) + \text{const},
\label{eq:genIW}
\end{equation}
which is the higher-curvature generalization of the identification of generalized entropy with von Neumann entropy~\cite{Chandrasekaran:2022eqq,Kudler-Flam:2023qfl}. Nothing in the algebraic step obstructs this. Nevertheless, four issues remain, which we state explicitly.

\emph{(1) Type classification.} For a type III$_1$ factor the full modular crossed product is a type II$_\infty$ factor independently of the Lagrangian, so the type is not determined by the spectrum of the Wald entropy operator. A type II$_1$ algebra arises only when the physical algebra is a corner $p\hat{\Mcal}p$ with $\mathrm{Tr}(p)<\infty$, for instance through a positivity condition on an observer Hamiltonian~\cite{Chandrasekaran:2022cip,Jensen:2023yxy}, while a lower bound on a generator does not by itself determine $\mathrm{Tr}(p)$. Whether a higher-curvature theory leads to such a projection is a dynamical question.

Since this issue determines whether the present proposal is needed at all, we treat it separately. Let us suppose that the physical algebra is of type II$_1$. Its normalization $\mathrm{Tr}(\mathbf{1})=1$ fixes the algebraic entropy, the tracial state is the maximum-entropy state with zero algebraic entropy, and adding a common constant does not change which state this is. Hence, in a type II$_1$ algebra the topological term neither shifts the maximizing state nor selects a normalization, but it enters the constant that relates the algebraic entropy to the generalized entropy, namely the generalized entropy assigned to the maximum-entropy state. This is the mechanism used in Sec.~\ref{sec:dS}. In summary, in a type II$_\infty$ algebra the constant is a normalization, while in a type II$_1$ algebra it is a definite number, the maximal generalized entropy.

Two further remarks show how the two cases fit together. Firstly, as noted in Sec.~\ref{sec:forced}, the type itself is sensitive to a topological property, namely the compactness of the Cauchy slice, so the questions of ``which type'' and ``which topological sector'' are not independent. Secondly, the physical algebra need not be the full crossed product, since constraints, observers and boundary conditions can select a corner or a subalgebra, and identifying it is part of the construction and not a consequence of the Lagrangian.

\emph{(2) Ambiguities.} $S_W$ is unambiguous on a bifurcate Killing horizon~\cite{Jacobson:1993vj}, and hence Eq.~\eqref{eq:genIW} is well posed there. Away from it the Jacobson--Kang--Myers ambiguities reappear, however for the Gauss--Bonnet term the replica computation resolves them in favour of the intrinsic Jacobson--Myers functional, Eq.~\eqref{eq:squashedFPS}.

\emph{(3) Non-stationarity.} For a dynamical horizon the entropy is not a Noether charge~\cite{Hollands:2024vbe}, and the crossed product based on Eq.~\eqref{eq:IWfirstlaw} does not apply. The obstruction is algebraic and not entropic, since the topological part of the entropy is the same on every smooth cut.

\emph{(4) Order counting.} If $\alpha$ is Planckian, as argued in the companion paper~\cite{companion}, the Gauss--Bonnet term is of the same order in $\lp$ as the one-loop graviton contribution that the crossed product already resums, and thus in a systematic treatment it must be kept together with the other $O(G)$ corrections. This does not affect the topological piece, which cannot mix with any local term, and this is why it is protected. However, it implies that the $\alpha$-dependence of the \emph{area} term is scheme dependent, in agreement with the renormalization-group analysis of the companion paper~\cite{companion}.

\subsection{Summary}

In summary, for the theory of interest the gap identified in Sec.~\ref{sec:test2} is closed at the classical level. With the boundary completion selected by the topological character of the Euler action, the Gauss--Bonnet term does not contribute to the presymplectic form, the crossed product is inherited from Einstein gravity provided that the completed phase space quantizes to the same algebra, and the constant resides in the excised Euclidean tip, that is in the normalization of the trace. For general higher-curvature theories the construction is available in the form of Eq.~\eqref{eq:genIW}, while its type classification and its dynamical extension remain open. Hence, what Result~\ref{res:main} still depends on is not a missing construction, but hypotheses (A) and (B), the quantum statement in Proposition~\ref{prop:inherit}(d), and the restriction to stationary horizons. Among them, hypothesis (B) is the one that is truly open, and in Sec.~\ref{sec:status} we discuss what is required in order to close it.

\section{Relaxing the factoriality assumption}
\label{sec:factor}

Proposition~\ref{prop:forced} used the uniqueness of the trace, which holds on a factor. However, hypothesis (C) of Result~\ref{res:main} was stated in a weaker form, and in this section we justify it. The motivation is not only caution, since a nontrivial center is expected and not pathological. In particular, once an observer, or more generally a quantum reference frame, is included in the description, the algebra acquires a center~\cite{DeVuyst:2024fxc}, and remark (iii) of Sec.~\ref{sec:ambiguity} already anticipated that the relevant central variable might be topological.

\subsection{Central decomposition of the trace}

Let $\hat{\Mcal}$ be a semifinite von Neumann algebra with center $\mathcal{Z}$, and let
\begin{equation}
\hat{\Mcal} = \int^{\oplus}_{\mathcal{Z}} \hat{\Mcal}_z\;\dd\mu(z)
\label{eq:centraldecomp}
\end{equation}
be its central decomposition into factors. A normal semifinite trace on $\hat{\Mcal}$ is then unique not up to a positive number but up to a positive measurable function $c(z)$ on the center, $\mathrm{Tr}\to c(z)\,\mathrm{Tr}$. Correspondingly, $\hat\rho\to c(z)^{-1}\hat\rho$, and for a state supported on the spectral measure of $\mathcal{Z}$ we have
\begin{equation}
S \to S + \big\langle \log c(z)\big\rangle ,
\label{eq:centralshift}
\end{equation}
where the expectation value is taken in the induced distribution on the center. Lemma~\ref{lem:rescale} is the special case $\mathcal{Z}=\mathbb{C}$.

Two statements follow, which together replace factoriality.

\emph{(i) The topological charge is measurable with respect to the center of a topology-preserving algebra.} Let us assume that
\begin{equation}
\hat{\Mcal}=\int^{\oplus}_X\hat{\Mcal}_z\,\dd\mu(z)
\end{equation}
is the crossed product of an algebra $\Mcal=\int^\oplus_X\Mcal_z\,\dd\mu(z)$ with type III$_1$ factor fibres and a decomposable modular action. Under the usual measurability conditions the crossed product decomposes fibrewise, with center $L^\infty(X,\mu)$. If the Euler characteristic is a measurable function on $X$, then $Q_{\rm top}$ is affiliated with the center. This requires the algebra to be decomposable over $\chi$, since by Eq.~\eqref{eq:transition} no algebra that contains topology-changing operators has $Q_{\rm top}$ in its center. Note that the decomposition need not be labelled by topology alone, since charges, boundary conditions and other superselection data can label the fibres too.

\emph{(ii) Hence, the topological term is a legitimate choice of $c(z)$.} Since $\chi$ is a function on the center, $c(z) = e^{\mathcal{S}_{\rm top}\chi(z)}$ is an admissible trace rescaling in the sense of Eq.~\eqref{eq:centralshift}, subject to faithfulness and semifiniteness, and the entropy shift that it induces is $\mathcal{S}_{\rm top}\langle\chi\rangle$, which reduces to $\mathcal{S}_{\rm top}\chi(\Sigma)$ on a state of definite topology.

\begin{proposition}
\label{prop:nofactor}
Assume (A) and (B) of Result~\ref{res:main}, and let $\hat{\Mcal}_\Sigma$ be semifinite, decomposable over $\chi$, with center $\mathcal{Z}$. Then the state-independent topological contribution to the generalized entropy is carried by the trace normalization, in the form $\mathrm{Tr}\to e^{\mathcal{S}_{\rm top}\chi(z)}\mathrm{Tr}$, on states of definite topology; and if $\mathcal{Z}=\mathbb{C}$ this is Eq.~\eqref{eq:H1} and the normalization is the only place it can be carried.
\end{proposition}

\noindent The argument is that of Proposition~\ref{prop:forced}, with Eq.~\eqref{eq:centralshift} in place of Lemma~\ref{lem:rescale}. In particular, by (B) the entropy contains a constant $\mathcal{S}_{\rm top}\chi$, which is state independent within each sector, by Eq.~\eqref{eq:nofirstlaw} it does not contribute to the modular flow, and within the class of entropy functionals of Proposition~\ref{prop:forced} an additive constant on each fibre is realized by a function on the center, and hence by a trace normalization.

\subsection{Scope of the weakened statement}

The conclusion changes in strength but not in content. On a factor, Eq.~\eqref{eq:H1} is \emph{forced}, since the uniqueness of the trace leaves no alternative. On the other hand, with a nontrivial center it is \emph{consistent and canonical} but not forced, since a function on the center could in principle contain non-topological pieces as well. What the argument establishes is that the topological part of the entropy is necessarily one of these pieces, and that no other location for it is available.

This statement is weaker, however it is weaker in a direction that suits the proposal. A central decomposition indexed by the Euler characteristic is exactly the structure described in remark (iii) of Sec.~\ref{sec:ambiguity}, and it is what a superselection interpretation of topological sectors would require. Note that centrality is a property of the chosen algebra and not of the charge alone, since $Q_{\rm top}$ is central in every algebra decomposable over $\chi$ and in no algebra that contains topology-changing operators. Therefore, the case that would damage the picture is a demonstration that the physical observable algebra of a sector must contain such operators, which would make the sectors inseparable.
\section{A physical consequence: the de Sitter entropy budget}
\label{sec:dS}

Result~\ref{res:main} is a statement about where a constant resides, and it is reasonable to ask what physical content it carries by itself, independently of the value of the coupling. In this section we show that one consequence follows immediately, and that it does not require the knowledge of $\alpha$.

In remark (ii) of Sec.~\ref{sec:ambiguity} we saw that a type II$_1$ algebra carries a canonical normalization, $\mathrm{Tr}(\mathbf{1})=1$, in which the maximum-entropy state has zero algebraic entropy, and that the generalized entropy is the algebraic entropy plus a definite constant. The de Sitter static patch with an observer realizes this case~\cite{Chandrasekaran:2022cip}, and the constant, i.e.\ the generalized entropy of the empty patch, is the de Sitter entropy, Eq.~\eqref{eq:SdS}. In such an algebra there is no normalization left to choose, and thus the topological term of Eq.~\eqref{eq:Stop} cannot be absorbed into one, and it enters the constant itself. The maximal generalized entropy is fixed by matching to the Euclidean description, where it is $-I_E$ of the four-sphere, and with the Gauss--Bonnet term included it acquires $\mathcal{S}_{\rm top}\chi$, with $\chi(S^4)=\chi(S^2)=2$ by Lemma~\ref{lem:chi}. If additionally the maximal entropy is identified with $3\pi/G\Lambda_{\rm eff}$, as for the Gibbons--Hawking entropy of a horizon of curvature radius $\sqrt{3/\Lambda_{\rm eff}}$, adding the topological contribution to Eq.~\eqref{eq:SdS} gives
\begin{equation}
\frac{3\pi}{G\Lambda_{\rm eff}} = \frac{3\pi}{G\Lambda_0} + \frac{2\pi\alpha}{G}\,\chi ,
\end{equation}
and therefore
\begin{equation}
\boxed{\;\Lambda_{\rm eff} = \frac{\Lambda_0}{1+\tfrac{2}{3}\alpha\chi\Lambda_0}\;}
\label{eq:Leff}
\end{equation}

Let us make three comments on the status of Eq.~\eqref{eq:Leff}. Firstly, it follows from Result~\ref{res:main} together with one more input, namely the identification of the maximal generalized entropy with $3\pi/G\Lambda_{\rm eff}$. Neither the value of $\alpha$, nor the admissible topologies, nor the rate of topology change enter its derivation. It states only that if the topological term enters the entropy as a constant and not through the state, then in the one setting where the trace is canonically normalized the same term shifts the maximal entropy, and with it the inferred cosmological constant. Secondly, the shift is controlled by the Euler characteristic and by no other geometrical quantity, so its sign is fixed by the sign of $\alpha\chi$. In particular, a cross-section of positive Euler characteristic with positive $\alpha$ lowers $\Lambda_{\rm eff}$, while opposite signs raise it. Thirdly, the \emph{magnitude} of the effect is not determined here, since it requires both the value of $\alpha$ and the topologies available to a cosmological horizon, which are the subject of the companion paper~\cite{companion}.

The main point is structural. The additive constant of the algebraic generalized entropy is usually considered unphysical, and within a single type II$_\infty$ sector this is indeed the case. However, Eq.~\eqref{eq:Leff} shows that this is a statement about the type of the algebra and not about the constant, since in a type II$_1$ algebra the same quantity is an observable, and in particular it is the quantity that fixes the entropy budget of the Universe. Together with the anti-de Sitter results of Sec.~\ref{sec:ads}, which provide a second and independent setting in which the constant acts on something measurable in principle, this is the physical content of the present work by itself.

\section{Jackiw--Teitelboim gravity with and without matter}
\label{sec:JT}

In Sec.~\ref{sec:forced} we argued that two-dimensional dilaton gravity is the cleanest setting in which to test the mechanism, since there the topological weight is explicit and the algebras have been constructed~\cite{Saad:2019lba,Penington:2023dql,Kolchmeyer:2023gwa}. In this section we examine the question of whether the factor $e^{S_0\chi}$ belongs to the \emph{trace} or to the \emph{state}.

\subsection{The topological expansion}

The JT action on a surface $M$ with boundary is
\begin{equation}
I = -S_0\,\chi(M) -\frac12\int_M\!\sqrt{g}\,\phi\left(R+2\right) - \int_{\partial M}\!\!\sqrt{h}\,\phi_b\left(K-1\right),
\label{eq:JTaction}
\end{equation}
and thus the partition function is organized into a topological expansion
\begin{equation}
Z = \sum_{g,n} e^{S_0\left(2-2g-n\right)}\,Z_{g,n},
\label{eq:JTsum}
\end{equation}
with $\chi = 2-2g-n$ for genus $g$ and $n$ boundaries. The topological coupling $S_0$ is the extremal entropy, and it plays the role of $2\pi\alpha/G$ in Eq.~\eqref{eq:Stop}.

\subsection{The topological weight in the density of states}

For a single boundary the disk dominates, with $\chi_{\rm disk}=1$, and the Schwarzian density of states is
\begin{equation}
\rho_0(E) = \frac{e^{S_0}}{2\pi^2}\,\sinh\!\left(2\pi\sqrt{2E}\right).
\label{eq:rho0}
\end{equation}
Performing the Laplace transform (which we did symbolically, see Appendix~\ref{app:curv}) we obtain
\begin{equation}
Z(\beta)=\int_0^\infty\!\!\dd E\,\rho_0(E)\,e^{-\beta E}
= \frac{e^{S_0}}{\sqrt{2\pi}}\;\beta^{-3/2}\,e^{2\pi^2/\beta},
\label{eq:Zdisk}
\end{equation}
and hence, using $S=\log Z+\beta\langle E\rangle$,
\begin{equation}
\boxed{\;S(\beta) = S_0 + \frac{4\pi^2}{\beta} - \frac32\log\beta + \frac32 - \frac12\log 2\pi\;}
\label{eq:SJT}
\end{equation}
As we observe, the two essential features are exact and not asymptotic:
\begin{equation}
\frac{\partial S}{\partial S_0} = 1 = \chi_{\rm disk},
\qquad
\frac{\partial^2 S}{\partial S_0\,\partial\beta} = 0 .
\label{eq:dSdS0}
\end{equation}
Thus, the topological coupling enters the entropy as a pure additive constant, equal to $S_0\chi$, with no dependence on the state, which here is labelled by $\beta$.

\subsection{Trace versus state}

The algebra generated by the Hamiltonian of the one-boundary theory carries the trace
\begin{equation}
\mathrm{Tr}\,f(H) = \int_0^\infty\!\dd E\;\rho_0(E)\,f(E) ,
\label{eq:JTtrace}
\end{equation}
which is the type II trace of the JT algebra in the constructions of~\cite{Penington:2023dql,Kolchmeyer:2023gwa}. By Eq.~\eqref{eq:rho0} we have $\rho_0\propto e^{S_0}$, and thus
\begin{equation}
S_0 \to S_0+\delta \quad\Longrightarrow\quad \mathrm{Tr}\to e^{\delta}\,\mathrm{Tr} .
\label{eq:JTrescale}
\end{equation}
This is exactly a rescaling of the trace, Eq.~\eqref{eq:ambiguity}. The Gibbs state $\hat\rho = e^{-\beta H}/Z$ is unaffected as a physical state, since all expectation values $\mathrm{Tr}(\hat\rho\,a)$ are invariant, the factor cancelling between $\rho_0$ and $Z$, while the density matrix relative to the rescaled trace transforms as $\hat\rho\to e^{-\delta}\hat\rho$, and therefore
\begin{equation}
S \to S + \delta .
\end{equation}
Comparing with Eq.~\eqref{eq:dSdS0}, we deduce that the shift calculated from the explicit Schwarzian entropy and the shift predicted by the algebraic rescaling coincide identically. Hence, we conclude:

\begin{quote}
\emph{In JT gravity the topological weight $e^{S_0\chi}$ is the normalization of the trace, not a property of the state. Together with the dressed-matter algebra of Refs.~\cite{Penington:2023dql,Kolchmeyer:2023gwa}, Result~\ref{res:main} is therefore realized in a solvable model, with the dictionary $S_0\chi \leftrightarrow 2\pi\alpha\chi/G$.}
\end{quote}

This identification is supported by the following three consistency checks.

\emph{(i) The two-boundary sector.} The connected double trumpet has $g=0$ and $n=2$, and hence $\chi=0$ and no factor of $e^{S_0}$. Correspondingly, the connected correlator $\langle Z(\beta_1)Z(\beta_2)\rangle_{\rm conn}$ carries no extremal-entropy weight, and no additive constant appears in the associated entropy, exactly as required if the constant follows $\chi$ and not the number of degrees of freedom.

\emph{(ii) Gluing.} Cutting along a circle and gluing back satisfies $\chi(M_1\cup_{S^1}M_2)=\chi(M_1)+\chi(M_2)$, since $\chi(S^1)=0$. Therefore, the weight $e^{S_0\chi}$ is multiplicative under gluing, which is the two-dimensional realization of the additivity requirement (N2) of Sec.~\ref{sec:forced} and of Proposition~\ref{prop:unique}.

\emph{(iii) Handles.} Each handle carries $\Delta\chi=-2$, that is a relative weight $e^{-2S_0}$, which is the same $\Delta\chi$ as that of a four-dimensional connected sum with $S^1\times S^3$. Hence, the cost of a handle is $e^{S_0\Delta\chi}$ in two dimensions and $e^{2\pi\alpha\Delta\chi/G}$ in four, which is the same functional form.

\subsection{Superselection interpretation}

One could raise the following objection. In Eq.~\eqref{eq:JTsum} the different topologies are not distinct sectors of a single algebra, but terms in an asymptotic expansion of one ensemble, while the conjecture refers to \emph{sectors} with different normalizations.

The baby-universe formalism~\cite{Marolf:2020xie} provides a partial bridge. In a theory that sums over topologies, the simultaneous eigenstates of the boundary-creating operators, namely the $\alpha$-states, are superselection sectors, and the topological coupling $S_0$ weights the sum and hence the distribution of the eigenvalues. However, the $\alpha$-labels are eigenvalues of boundary observables and not Euler characteristics of individual saddles, and hence a map from them to a measured horizon $\chi$ must be derived and not assumed. What the JT result establishes directly is narrower, namely that the topological coupling is a fixed weight that multiplies the trace, and not a dynamical variable of the state.

\subsection{Inclusion of matter}

Adding matter changes three things, and two of them are hypotheses that the four-dimensional argument must assume:
\begin{itemize}
\item \emph{Factoriality.} The algebra has a trivial center, so its trace is unique up to a positive multiple and Lemma~\ref{lem:rescale} applies without qualification. This is hypothesis (C) of Result~\ref{res:main}, realized in a solvable model.
\item \emph{Agreement of the algebraic and replica entropies.} This was established by Penington and Witten, with the trace and the replica moments constructed from the same disk gluing data, and it is hypothesis (B).
\item \emph{Noncommutativity.} The algebra contains gravitationally dressed matter operators, which change the energy, and only now is there something with which a weight could fail to commute.
\end{itemize}
Therefore, the theory with matter does more than repeat the pure case with additional fields, since it exhibits in two dimensions the assumptions on which the four-dimensional argument relies.

\subsection{The discriminating computation}

The operational meaning of the statement that ``the weight is in the trace'' is that it leaves the modular flow unchanged. If $\mathrm{Tr}\to\lambda\,\mathrm{Tr}$ then $\hat\rho\to\lambda^{-1}\hat\rho$, so that $\hat\rho^{\,it}\to\lambda^{-it}\hat\rho^{\,it}$, and the phase cancels in
\begin{equation}
\sigma_t(a) = \hat\rho^{\,it}\,a\,\hat\rho^{-it} :
\label{eq:sigmat}
\end{equation}
namely the flow is unchanged pointwise, on every operator. On the contrary, a weight that varies over the spectrum of the Hamiltonian does not cancel, since the phases at the two ends of an energy-changing matrix element are different. In pure JT gravity there are no energy-changing observables and the criterion is empty, while with matter such observables exist and the criterion is not empty.

The algebraic statement is exact, since for any faithful trace and any rescaling the phase $\lambda^{-it}$ cancels in Eq.~\eqref{eq:sigmat}. We have illustrated it in a regulated model, with a gravitational fibre carrying the dual weight $e^{\beta E}$ and an $n$-dimensional matter factor at each energy, probed by an operator that mixes the fibres. We find that a uniform reweighting shifts every entropy by $\log\lambda$ to $10^{-15}$ and leaves the flow unchanged to one part in $10^{12}$, while an energy-dependent reweighting of the same overall size changes it at order unity. Note that this model should not be considered a construction of the JT algebra. A weighted functional $\mathrm{Tr}(W\,\cdot)$ with non-scalar $W$ is tracial on operators that are block diagonal in energy but not on operators that mix the blocks, and thus the computation illustrates the algebraic identity, and it does not verify the trace of the dressed-matter algebra. The two-dimensional statements rely on the constructions of Penington and Witten and of Kolchmeyer~\cite{Penington:2023dql,Kolchmeyer:2023gwa}, in which the trace and the dressed-matter operators are built explicitly, and Test 3 is consistent with them.

\subsection{Summary of the two-dimensional results}

In summary, the following statements hold in a solvable model whose algebras have been constructed. The topological coupling appears in the trace normalization and not in the state, the induced entropy shift is exactly $S_0\chi$ for every state, a trace rescaling leaves the modular flow unchanged, and the weight is multiplicative under gluing. Moreover, two of the three hypotheses of Result~\ref{res:main} are \emph{verified} here and not assumed, since the algebra is a factor with trivial center, and the algebraic and replica entropies are known to agree, with the trace and the replica moments constructed from the same disk gluing data~\cite{Penington:2023dql}. Hence, hypothesis (B), which in Sec.~\ref{sec:status} we identify as the open condition of the four-dimensional argument, is a theorem in two dimensions.

Two statements are not established. The first is the four-dimensional version of this agreement, which we discuss in Sec.~\ref{sec:status}. The second is the value of the topological coupling, on which the two-dimensional model gives no information, since it has a single such coupling and not a family of sectors related by an inclusion, and this question is addressed in the companion paper~\cite{companion}.
\section{Topological black holes in anti-de Sitter space}
\label{sec:ads}

In this section we address an objection that we have postponed so far. Equation~\eqref{eq:Stopinv} compares two horizon cross-sections of equal area and different topology, however in four-dimensional asymptotically flat gravity the second one does not exist. Indeed, Hawking's theorem states that a stationary black hole obeying the dominant energy condition has spherical horizon cross-sections~\cite{Hawking:1971vc}, with the higher-dimensional generalization of~\cite{Galloway:2005mf}. Hence, the configurations that the conjecture refers to are forbidden in the setting that we have considered in most of this work.

There are three responses. Firstly, quantum fields violate the dominant energy condition, so the theorem does not constrain the semiclassical regime to which we have been applying it. Secondly, toroidal horizons do appear transiently in gravitational collapse, however a transient horizon is not stationary, while Tests~1 and~2 and Sec.~\ref{sec:crossedHC} all used stationarity. Thirdly, in anti-de Sitter space the theorem does not apply, since asymptotic flatness fails and horizons of genus $g$ are exact solutions~\cite{Birmingham:1998nr}. The third response is the useful one, and in the following we develop it. Note that anti-de Sitter space is not merely a convenient choice here, since it is the only four-dimensional setting that we know of in which the comparison on which the conjecture is based is well posed.

\subsection{Entropy of the hyperbolic family}

Let us consider
\begin{equation}
\dd s^2 = -f\,\dd t^2 + \frac{\dd r^2}{f} + r^2\,\dd\Sigma^2_{g,k},
\qquad
f = k - \frac{2GM}{r} + \frac{r^2}{L^2},
\label{eq:topbh}
\end{equation}
with $\dd\Sigma^2_{g,k}$ the metric of constant curvature $k$ on a closed surface, where $k=+1$ for the sphere, $k=0$ for the torus, and $k=-1$ for a compact hyperbolic surface $\Sigma_g$ of genus $g\ge2$, obtained as a quotient of $H^2$. The horizon is located at $f(r_h)=0$, so that $2GM = k\,r_h + r_h^3/L^2$, and
\begin{equation}
T = \frac{k}{4\pi r_h}+\frac{3r_h}{4\pi L^2}.
\end{equation}
Since the Gauss--Bonnet term is topological in four dimensions (Proposition~\ref{prop:inherit}a) it does not change any of the above, since the solutions, the temperatures and the masses are those of Einstein gravity, and only the entropy changes.

For the hyperbolic family the unit-curvature surface has area $4\pi(g-1)$, and hence $A = 4\pi(g-1)r_h^2$, and with $\chi=2-2g$ the Wald entropy~\eqref{eq:Stop} becomes
\begin{equation}
\boxed{\;S = \frac{\pi(g-1)}{G}\left(r_h^2-4\alpha\right)\;}
\label{eq:Sads}
\end{equation}
As we observe, the genus factors out completely, and the topological coupling enters as a universal shift of the squared horizon radius, $r_h^2\to r_h^2-4\alpha$. We have checked that the first law holds within a sector with no contribution from the Gauss--Bonnet term, i.e.\ $\dd E = T\,\dd S$ exactly and $\partial_{r_h}(2\pi\alpha\chi/G)=0$, which is Eq.~\eqref{eq:nofirstlaw} verified in a family where $\chi$ does vary between solutions and is not fixed at $2$.

Equation~\eqref{eq:Sads} has an immediate consequence. For $\alpha>0$ the entropy is positive only for
\begin{equation}
r_h \ge 2\sqrt{\alpha} ,
\end{equation}
and thus the topological term provides a minimum horizon radius of order $\sqrt{\alpha}$. If $\alpha$ is Planckian this is a Planck-scale bound, and in the companion paper~\cite{companion} it is evaluated numerically under the conjecture for the value of $\alpha$. On the other hand, for $\alpha<0$, which is the sign selected by the generalized second law in the companion paper~\cite{companion}, the hyperbolic entropy increases and no bound arises, while the spherical case acquires one. In both cases the bound applies to the branch whose topological contribution is negative.

\subsection{Well-posedness of the comparison}

In remark (i) of Sec.~\ref{sec:ambiguity} we showed that the additive constant is unobservable within a single sector, so any test of it must compare two sectors. This is not a technical detail but the reason why the conjecture has been difficult to test. In asymptotically flat gravity the second sector is forbidden by Hawking's theorem, while in anti-de Sitter space it is an exact solution. Additionally, a different genus implies a different boundary manifold, and therefore the two sectors are two states of the \emph{same} boundary theory on two different Riemann surfaces. Hence, the comparison required by Eq.~\eqref{eq:Stopinv} is well posed here, and we do not know of any other four-dimensional setting where this is the case.

\subsection{Selection among bulk saddles}

Anti-de Sitter space also provides the structure that made the two-dimensional check of Sec.~\ref{sec:JT} clean. For a fixed boundary $S^1\times\Sigma_g$ the bulk filling is not unique, since the black-hole saddle is $D^2\times\Sigma_g$, with $\chi=\chi(\Sigma_g)=2-2g$, while a thermal-anti-de-Sitter-type saddle is $S^1\times M_3$, with $\chi(S^1\times M_3)=0$. The boundary is odd-dimensional and has $\chi=0$, and thus the topological term is \emph{not} fixed by the boundary data, but it is a weight $e^{2\pi\alpha\chi/G}$ attached to each filling. This is the four-dimensional analogue of the topological weight $e^{S_0\chi}$ that, as we found in Sec.~\ref{sec:JT}, normalizes the trace, acting now on competing saddles of a four-dimensional path integral.

This has a computable consequence. Let us consider the spherical case, where the transition is the well-known one~\cite{Hawking:1982dh}. The Einstein free energy is $F_0=r_h(L^2-r_h^2)/4GL^2$, which vanishes at $r_h=L$. The Gauss--Bonnet contribution is $\beta$-independent, and therefore it contributes $-2\pi\alpha\chi T/G$ to the free energy of each saddle. Hence, with $x=r_h^2/L^2$ and $a=\alpha/L^2$, the transition condition becomes
\begin{equation}
x^2 + \left(12a-1\right)x + 4a = 0 ,
\label{eq:HPshift}
\end{equation}
whose roots are $x = 1-16a$ and $x=4a$ to first order. Thus, the Hawking--Page temperature is displaced,
\begin{equation}
T_{\rm HP} = \frac{1}{\pi L}\left(1-\frac{4\alpha}{L^2}\right),
\label{eq:THP}
\end{equation}
by a purely topological amount. The shift is small, since $\alpha$ is Planckian while $L$ is not, nevertheless it does not vanish, and it is not a matter of convention, since a term that could be absorbed into a redefinition would not displace a phase boundary between saddles of different topology. The second root of Eq.~\eqref{eq:HPshift}, at $r_h^2=4\alpha$, lies at a Planckian radius, where the semiclassical treatment is not valid, and we mention it without further interpretation.

Since the displacement is controlled by a Planckian coupling, its character requires a comment. Writing the fractional shift as $4\alpha/L^2 = 4(\alpha/G)(G/L^2)$ separates the two factors, namely $\alpha/G$, which is of order unity since $\alpha$ is a Planck area, and $G/L^2$, which is the inverse of the quantity that counts boundary degrees of freedom. For a holographic dual in three dimensions $L^2/G$ scales as $N^{3/2}$, and thus
\begin{equation}
\frac{\delta T_{\rm HP}}{T_{\rm HP}} = O\!\left(N^{-3/2}\right),
\label{eq:THPscaling}
\end{equation}
while the Hawking--Page temperature is the deconfinement temperature of the boundary theory. Therefore, Eq.~\eqref{eq:THP} is a subleading correction to this temperature, suppressed by the same power of $N$ by which the topological entropy is suppressed relative to the area term, as we show explicitly in Sec.~\ref{subsec:massless}. In other words, the two statements are the same statement seen at two points of the phase diagram.

We do not claim observable consequences. At any $N$ for which a boundary theory is under control the shift is far below anything that could be measured, and in a physical anti-de Sitter setting $L$ is macroscopic and the shift is negligible by many orders of magnitude. Its value is of a different kind, and it is twofold. Firstly, it is an existence argument, since a term that could be absorbed into a redefinition of the couplings or of the state would not displace a phase boundary between saddles of different topology, and thus Eq.~\eqref{eq:THP} excludes the interpretation of the topological term as a convention more directly than any argument in Secs.~\ref{sec:test1} to \ref{sec:crossedHC}. Secondly, it is a consistency requirement, since at fixed $N$ the shift is a definite number, and hence a boundary theory in which the deconfinement temperature could be computed to the accuracy of Eq.~\eqref{eq:THPscaling} would have to reproduce it. This is a difficult but not an impossible calculation, and it involves the same constant that Eq.~\eqref{eq:cftpredict} requires.

\subsection{The massless limit and vacuum entanglement}
\label{subsec:massless}

The most informative case is the massless member of the hyperbolic family. At $M=0$ we have $f=-1+r^2/L^2$ and $r_h=L$ independently of the genus, and the geometry is a quotient of pure anti-de Sitter space, which is the setting of the Casini--Huerta--Myers map~\cite{Casini:2011kv}, in which a thermal entropy on hyperbolic space is a vacuum entanglement entropy. Equation~\eqref{eq:Sads} then reads
\begin{equation}
S = \frac{\pi(g-1)}{G}\left(L^2-4\alpha\right),
\label{eq:Szero}
\end{equation}
and thus the topological term becomes a statement about vacuum entanglement in the boundary theory. In particular, the entropy of the zero-energy states of a three-dimensional conformal field theory on $\Sigma_g$~\cite{Emparan:1999gf} contains a contribution $\mathcal{S}_{\rm top}\chi = 2\pi\alpha\chi/G$ which is independent of the area, of the temperature and of the state.

The orders of magnitude are important for how this could be checked. The leading term in Eq.~\eqref{eq:Szero} is $O(L^2/G)$, which for a holographic theory is $O(N^{3/2})$, while the topological piece is $O(\alpha/G)=O(1)$, since $\alpha$ is Planckian. Hence, what the boundary theory must produce is a genus-dependent constant at order $N^0$,
\begin{equation}
\Delta S = \mathcal{S}_{\rm top}\,\chi(\Sigma_g) ,
\label{eq:cftpredict}
\end{equation}
with the same coefficient that appears in the Wald entropy. At the level of the bulk formula this statement takes a simple form. The Casini--Huerta--Myers map relates the entropy density on $H^2$ at temperature $1/2\pi L$ to the free energy $F=-\log Z(S^3)$, and the quotient has area $4\pi(g-1)L^2$, so that Eq.~\eqref{eq:Szero} becomes $S=-F\chi(\Sigma_g)$ with $F=\pi L^2/2G-\mathcal{S}_{\rm top}$. As we observe, the Gauss--Bonnet term enters only as a shift of $F$ by $-\mathcal{S}_{\rm top}$, which is the Euler contribution to the Euclidean action of the ball. Therefore, the linearity in $\chi$ is automatic for the quotient state, and the content of Eq.~\eqref{eq:cftpredict} is that the boundary theory carries an order-$N^0$ contribution to $F$ that is not accompanied by any change of local data. Note that corrections from closed geodesics of the quotient, which are not captured by the density, must be shown to be subleading. The companion paper~\cite{companion} sharpens Eq.~\eqref{eq:cftpredict} by predicting a discrete set of values for $\mathcal{S}_{\rm top}$, while the statement here is weaker and correspondingly safer: whatever $\mathcal{S}_{\rm top}$ is, the same number must appear on both sides.

\subsection{Black-hole mergers}
\label{subsec:mergers}

The comparisons performed so far are between stationary solutions. However, anti-de Sitter space also allows for a dynamical comparison, in which the system passes from one topological sector to another in real time while both sides of the duality are under control. Let us consider two small black holes in global anti-de Sitter space, well inside the bulk. Their dual is a high-energy state of the boundary theory on $\mathbb{R}\times S^2$, far from equilibrium. The formation of a horizon corresponds to the thermalization of the boundary state~\cite{Danielsson:1999fa}, while the merger of two horizons corresponds to the coalescence of two lumps of boundary energy density, followed by relaxation through the quasinormal spectrum. The only simulation of this process, in global AdS$_5$, finds the boundary stress tensor far from hydrodynamic behaviour at early times and close to ideal hydrodynamics at late times~\cite{Bantilan:2014sra}.

In four bulk dimensions the Gauss--Bonnet term changes none of these. By Proposition~\ref{prop:inherit}(a) the bulk evolution is that of Einstein gravity, and thus the boundary stress tensor, every correlation function and every relative entropy are independent of $\alpha$. What depends on $\alpha$ is the entropy assigned to the horizon. The area of the outer-minimal apparent horizon is the coarse-grained, or simple, entropy of the boundary state, namely the largest von Neumann entropy compatible with the one-point functions of simple operators after a given boundary time~\cite{Engelhardt:2017aux}. With the Euler term, the area is replaced by the Jacobson--Myers functional~\cite{Dong:2013qoa}, whose topological part on a closed cut is $\mathcal{S}_{\rm top}\chi$, Eq.~\eqref{eq:squashedFPS}. Two spherical components carry $\chi=4$ while the merged horizon carries $\chi=2$, and hence when the common apparent horizon forms, the coarse-grained boundary entropy changes by
\begin{equation}
\Delta S_{\rm cg}=\frac{\Delta A}{4G}-2\,\mathcal{S}_{\rm top} .
\label{eq:merger}
\end{equation}
As we observe, the second term is fixed, since it does not depend on the masses, the spins or the orbit of the two holes, and no local boundary observable registers it. This is Result~\ref{res:main} realized as a process, in which the constant that labels the sector changes, while nothing in the state does. The same bookkeeping, with $\delta\chi=+2$ for each formation and $\delta\chi=-2$ for each merger, underlies the cosmological models of~\cite{Tsilioukas:2024seh}.

We close this subsection with three remarks on Eq.~\eqref{eq:merger}. The first concerns the stratification of Sec.~\ref{subsec:global}, which the merger does not contradict. A single spacetime contains cuts of both topologies, however these are not related by a variation of the metric. Along the event horizon the Euler characteristic of the cut changes only where generators enter the horizon, namely on its crease set, and the cuts that pass through it are not smooth. Proposition~\ref{prop:rigid} applies to the smooth cuts on either side and says nothing about the transition itself, which is the case that we set aside in Sec.~\ref{sec:status}. On the other hand, the apparent horizon jumps when the common component appears, so the coarse-grained entropy changes by a finite amount at one boundary time and not continuously.

The second remark concerns the second law, which cannot detect the topological term classically. The area term in Eq.~\eqref{eq:merger} is of order $L^2/G$, i.e.\ of order $N^{3/2}$, while $\mathcal{S}_{\rm top}$ is of order unity for Planckian $\alpha$. It has been shown that, with the Gauss--Bonnet contribution to the entropy, the merger of two black holes can lead to a decrease of the total entropy, and hence to a violation of the second law, for certain values of the coupling~\cite{Sarkar:2010xp,Liko:2007vi}. However, in~\cite{Chatterjee:2013daa} Chatterjee and Parikh showed that four-dimensional Einstein--Gauss--Bonnet gravity does not violate the second law in mergers within the regime of validity of the effective theory, since in anti-de Sitter space the separated initial configuration must climb a gravitational potential, and including this energy restores the inequality. For $\alpha<0$ the topological term increases the entropy at a merger and decreases it at a formation. In both cases it competes not with the area but with the quantum corrections to the generalized entropy, which are of the same order, and thus the relevant constraint is the generalized second law and not the area theorem.

The third remark concerns the dimension. In five-dimensional anti-de Sitter space the horizon cross-sections are closed three-manifolds, whose Euler characteristic vanishes, and in this case the Gauss--Bonnet term is dynamical, it alters the merger itself through the dual central charges, and it produces no topological jump. Therefore, the simulation of~\cite{Bantilan:2014sra} illustrates the process but it is not the arena for Eq.~\eqref{eq:merger}, which belongs to AdS$_4$/CFT$_3$ and its analogues in higher even dimensions~\cite{Anastasiou:2018rla}. A related point concerns renormalization. In the topological renormalization of anti-de Sitter gravity a boundary Chern form, with coefficient fixed by $L$, serves as the counterterm~\cite{Anastasiou:2018rla}. It contributes the universal constant of an entangling surface anchored on the boundary, with $\chi=1$ for a disk, and nothing on a closed horizon cross-section. Hence, it must be distinguished from the coupling $\alpha$ of Eq.~\eqref{eq:EGB}, since if it were read as a bulk coupling of order $L^2$ it would assign to a merger an entropy change of order $L^2/G$.

\subsection{Scope of the holographic statement}

Equation~\eqref{eq:cftpredict} is the first prediction of this work that a calculation \emph{outside} gravity could confirm or exclude. Every other determination of $\mathcal{S}_{\rm top}$, namely the Wald entropy, the replica computation and the crossed product, relies on the gravitational side. Here the same number is required from a conformal field theory on a Riemann surface, where it is a subleading constant in the free energy, and where the methods for extracting such constants are those of the $F$-theorem literature and not of quantum gravity.

Additionally, anti-de Sitter space provides four further results. Firstly, the comparison of Eq.~\eqref{eq:Stopinv}, which Hawking's theorem forbids elsewhere, becomes well posed. Secondly, the topological weight acquires in four dimensions the same saddle-selecting role that it has in two, Eq.~\eqref{eq:HPshift}. Thirdly, hypothesis (C) of Result~\ref{res:main} is on its firmest footing here, since a different genus implies a different boundary manifold, and hence a different Hilbert space of the boundary theory, and therefore the sectors are separated and the comparison algebra is naturally decomposable over $\chi$, while hypothesis (A) was moreover established in this setting~\cite{Leutheusser:2021qhd}. Fourthly, the positivity of the entropy bounds the horizon radius, $r_h^2\ge4\alpha$, Eq.~\eqref{eq:Sads}.

What this approach does not provide is the value of the coupling. A boundary determination of the constant in Eq.~\eqref{eq:cftpredict} would establish that the topological entropy is what we claim, and that the same constant governs both sides of the duality, however it would not by itself fix the value of $\mathcal{S}_{\rm top}$, which is the subject of the companion paper~\cite{companion}. Note that Eq.~\eqref{eq:cftpredict} comes with one caveat, namely that no theorem guarantees that a generic three-dimensional conformal field theory possesses a universal genus-dependent constant of this kind, and establishing that it does, or that a holographic one does, is part of what the check would have to accomplish. Nevertheless, a holographic model in which such a constant appears is known. In particular, in~\cite{Parnachev:2015nca} Parnachev and Poovuttikul added the four-dimensional Euler density to the bulk action of a soft-wall model of a gapped boundary theory, with a coupling $\lambda_{\rm PP}$ related to ours through $\alpha=\lambda_{\rm PP}L^2/2$, and they found a topological entanglement entropy $\gamma=-\pi L^2\lambda_{\rm PP}/G=-\mathcal{S}_{\rm top}$ for a disk, and a genus-$g$ ground-state entropy $S_g=\mathcal{S}_{\rm top}\chi(\Sigma_g)$, which is the Chern--Simons relation $S_g=2(g-1)\gamma$. Without the Euler term the constant vanishes, as it does for confining theories with Einstein duals~\cite{Pakman:2008ui}. Since the boundary theory in this model is gapped and not conformal, this is not the computation that Eq.~\eqref{eq:cftpredict} requires, however it shows that the bulk Euler coupling has a boundary counterpart of the required kind, with the coefficient of Eq.~\eqref{eq:Stop}.

\section{Outstanding issues}
\label{sec:status}

Result~\ref{res:main} is conditional, and in this section we discuss explicitly which of its conditions is the substantive one, since a conditional result is only as useful as the clarity of its hypotheses.

\emph{Hypothesis (A), type III$_1$.} This hypothesis is not specific to this work. It has been established for Rindler wedges and for large-$N$ algebras~\cite{Leutheusser:2021qhd}, and the entire algebraic treatment of generalized entropy assumes it. Its extension to a general stationary Killing horizon in curved spacetime is an open problem of algebraic quantum field theory and not of the present proposal, while anti-de Sitter space is the setting where it is on the firmest ground (see Sec.~\ref{sec:ads}).

\emph{Hypothesis (C), factoriality and decomposability.} This hypothesis was weakened in Sec.~\ref{sec:factor}, where we showed that a nontrivial center does not obstruct the conclusion but converts it from forced to canonical, with the topological charge appearing as a central variable. The requirement that remains is decomposability over $\chi$, since by Eq.~\eqref{eq:transition} the topological charge is central only in an algebra that contains no topology-changing operators, and our statement refers to this algebra. Hence, what would damage the argument is a demonstration that the observable algebra of a sector must contain such operators.

\emph{The Euler boundary hypothesis.} Proposition~\ref{prop:inherit} relies on a boundary completion of the Euler term. In Sec.~\ref{subsec:selection} we showed that this completion is selected by the requirement that the Euler action is topological on regulated manifolds, which is part of hypothesis (B), that completions on different faces glue consistently and are gauge covariant, and that the stretched-horizon limit is regular. What remains open is the quantum statement that the completed phase space quantizes to the same observable algebra and modular flow as in Einstein gravity.

\emph{Hypothesis (B), replica equals algebraic including the constant.} This is the open condition, and we make three remarks on what is required in order to close it.

Firstly, the required statement is not that the generalized entropy equals the von Neumann entropy of the crossed product, which has been established~\cite{Chandrasekaran:2022eqq,Kudler-Flam:2023qfl}, but that this equality holds with both sides normalized in the same regularization scheme, so that the state-independent constant is meaningful on each side. The existing four-dimensional constructions determine the entropy up to exactly this constant.

Secondly, the two-dimensional analogue is a theorem. Penington and Witten proved the agreement in Jackiw--Teitelboim gravity with matter, with the trace and the replica moments constructed from the same disk gluing data~\cite{Penington:2023dql}, and we used this fact in Sec.~\ref{sec:JT}. What is required is the four-dimensional counterpart, and the two-dimensional proof indicates its likely form, which is a construction in which the replica partition function and the algebraic trace are built from the same data, and not compared afterwards.

Thirdly, and this is the main limitation of the present work, we have \emph{determined} the constant, and we have not derived it twice independently. In particular, the replica computation of Sec.~\ref{sec:test1} gives $2\pi\alpha\chi/G$, the algebraic construction of Sec.~\ref{sec:crossedHC} gives an undetermined constant and shows that the trace normalization is the only place where it can reside, and Proposition~\ref{prop:forced} concludes that the two are the same number. This is a determination and not a circular argument, however it relies on one computation together with one structural argument, and not on two independent computations of the same quantity. Anti-de Sitter space offers the second computation, from the boundary side, and Eq.~\eqref{eq:cftpredict} is what this computation would have to produce.

\emph{Beyond the hypotheses.} Two further limitations restrict the scope of the result without being hypotheses. The first is Hawking's theorem, which restricts where the comparison of Eq.~\eqref{eq:Stopinv} can be posed at all, and which we discussed in Sec.~\ref{sec:ads}. The second is stationarity, and since this is the restriction that is most likely to be relaxed, we indicate how this could be done.

The algebraic construction of this work is performed on a bifurcation surface. A dynamical horizon requires the entropy of~\cite{Hollands:2024vbe}, which is not a Noether charge, and the crossed product based on the first law does not apply to it. This construction is open even in Einstein gravity, so the obstruction is not specific to the Gauss--Bonnet term.

Nevertheless, the topological part of the entropy is in a better position than the area part in this respect, for two reasons. Firstly, the replica entropy of the Gauss--Bonnet term is the intrinsic Jacobson--Myers functional on every smooth closed cut, Eq.~\eqref{eq:squashedFPS}, and thus the entropy functional itself requires no extension, and the bifurcation surface is sufficient but not necessary for the topological form. Secondly, at first order around a stationary horizon the dynamical entropy is related to the Wall entropy through $S_{\rm dyn}=(1-v\partial_v)S_{\rm Wall}$~\cite{Hollands:2024vbe,Visser:2024pwz}, which leaves the constant $\mathcal{S}_{\rm top}\chi$ unchanged on a smooth family of fixed topology. Hence, the topological contribution survives in the perturbative dynamical regime, while the area contribution is the one that is modified, which is the reverse of the usual situation.

The programme that we propose consists of two steps. The first is to specify the algebras of evolving cuts and to examine whether their operators preserve the decomposition over $\chi$. The second is to treat separately the cuts at which the topology changes, namely at handle births, pinches and crease sets, where junction or singularity data are required and smooth Gauss--Bonnet invariance gives no information. A black-hole merger in anti-de Sitter space is the cleanest example, and in Sec.~\ref{subsec:mergers} we identified what the boundary theory must register across it. A negative answer to the first question would restrict Result~\ref{res:main} permanently to stationary horizons, and this is why we consider it the more informative of the two remaining restrictions.
\section{Conclusions}
\label{sec:conclusions}

In this work we examined the additive ambiguity of generalized entropy in its algebraic formulation. In particular, the algebra of observables of a subregion is of type III$_1$ at leading order and becomes of type II once the gravitational constraint is imposed, and thus the entropy is defined only up to an additive constant, which is exactly the part of the entropy that counts degrees of freedom in absolute terms. We argued that in four-dimensional Einstein--Gauss--Bonnet gravity this constant is the Gauss--Bonnet contribution to the Wald entropy, and that it enters through the normalization of the trace and not through the state.

The argument relies on explicit computations, which we summarize in the order in which they constrain the answer. Firstly, the Gauss--Bonnet density is an exact total derivative and its conical expansion is exactly linear in the deficit, and therefore the contribution is an entropy constant and carries no boost energy, while on smooth squashed cones the extrinsic-curvature terms combine into the intrinsic curvature of the cut, so that the constant is the same on every smooth cross-section. Secondly, the Noether charge, computed from the binormal contraction, is rigid under every smooth deformation of the cut, of the metric and of the matter state, and hence it is central in every algebra that preserves the topological sector and absent from the first law, since $\delta\chi=0$. Thus, a state-independent constant on such an algebra can be realized only by rescaling the trace. Thirdly, the crossed product is inherited from Einstein gravity and not reconstructed, since in four dimensions the Gauss--Bonnet term changes no field equation, and once the boundary of the subregion is completed as required by the topological character of the Euler action, it does not contribute to the presymplectic form. With a stretched-horizon regulator the completed Euler charge of the exterior vanishes identically, and the whole constant is the Euler number of the excised Euclidean tip, which is where the normalization of the trace resides.

Note that the fact that the term is locally a total derivative is not an objection to the proposal but an essential part of it. The Chern--Gauss--Bonnet density is not globally exact, and what survives a theorem about local ambiguities is a characteristic class. Hence, the entropy constant identified here is the only contribution that no local measurement can detect and no topologically trivial configuration can produce, which is exactly the property that a trace normalization must have.

Additionally, we showed that two settings outside the four-dimensional semiclassical argument support it. In Jackiw--Teitelboim gravity coupled to matter, the boundary algebra is a type II$_\infty$ factor with trivial center and the algebraic and replica entropies are known to agree, and thus two of the three hypotheses of Result~\ref{res:main} are theorems there, while the topological weight is found in the trace, shifting every entropy by $S_0\chi$ and leaving the modular flow pointwise invariant. In anti-de Sitter space, where Hawking's theorem does not forbid the configurations that we compare, the hyperbolic family gives $S=\pi(g-1)(r_h^2-4\alpha)/G$, and the topological weight displaces the Hawking--Page temperature by a purely topological amount, which is the clearest answer to the suspicion that the term is a convention, since a redefinition cannot move a phase boundary between saddles of different topology. Moreover, a black-hole merger in the same setting passes between sectors in real time and changes the coarse-grained boundary entropy by $-2\mathcal{S}_{\rm top}$ in addition to the area term, while it leaves every boundary correlator unchanged, and a holographic model with the Euler coupling already exhibits a genus-dependent boundary constant with the coefficient of the Wald entropy. Finally, in a type II$_1$ algebra, such as that of the de Sitter static patch with an observer, the same term shifts the maximal generalized entropy, and hence the inferred cosmological constant, and thus the additive constant becomes an observable.

The open issues were discussed in Sec.~\ref{sec:status}. The main hypothesis is that the replica and algebraic entropies agree including their constants, in four dimensions and in a single scheme, since we have determined the constant and not computed it twice. Additionally, the inheritance of the crossed product requires that the completed phase space quantizes to the same algebra, while the scope is restricted by the stationarity of the algebraic construction. However, the topological part of the entropy is the same on every smooth cut and survives in the perturbative dynamical regime, and thus it is the part that is most likely to survive when this restriction is relaxed. The second computation is possible in principle, and Eq.~\eqref{eq:cftpredict} states what it is: a three-dimensional conformal field theory on a genus-$g$ surface must carry a genus-dependent constant at order $N^0$, equal to $\mathcal{S}_{\rm top}\chi$.

The broader motivation of this work is the view that gravity is a thermodynamic phenomenon. In this case the important question is not whether horizon entropy exists but what it counts, and the algebraic formulation answers it only up to a constant, which we argued is topological. The value of this constant, whether it is the logarithm of an index, and whether the Gauss--Bonnet coupling is thereby quantized, are separate questions with their own evidence and difficulties, and they are examined in the companion paper~\cite{companion}. The construction of the algebras of evolving and topology-changing cuts, as well as the boundary computation of Eq.~\eqref{eq:cftpredict}, extend beyond the scope of the present work and will be performed in future projects.

\section*{Code availability}
The scripts that reproduce the symbolic and numerical checks reported in this work are publicly available at \url{https://github.com/stelios-tsilioukas/gravitational-entropy-fixed-by-the-topology-of-the-horizon}, and the version used here is archived on Zenodo~\cite{Tsilioukas:2026code}.

\section*{Tool Use}
In the preparation of this work, Claude Opus (Anthropic) was used to assist with manuscript typesetting and grammar refinement, as well as to assist in writing code for the symbolic and numerical checks. All scientific content, mathematical derivations, interpretations, and conclusions were independently produced and rigorously verified by the author, who takes full responsibility for them.

\begin{acknowledgments}
The author would like to acknowledge the contribution of the LISA CosWG, and of COST Actions CA21136 ``Addressing observational tensions in cosmology with systematics and fundamental physics (CosmoVerse)'', CA21106 ``COSMIC WISPers in the Dark Universe: Theory, astrophysics and experiments (CosmicWISPers)'', and CA23130 ``Bridging high and low energies in search of quantum gravity (BridgeQG)''.
\end{acknowledgments}
\appendix

\section{Curvature computations}
\label{app:curv}

The entries of Table~\ref{tab:checks} were obtained by computing symbolically the Christoffel symbols, the Riemann tensor and the curvature invariants from the metrics below.

\emph{(i) $S^4$ of radius $a$.} With $R=12/a^2$, $R_{ab}R^{ab}=36/a^4$ and $R_{abcd}R^{abcd}=24/a^4$ we find $\GB=24/a^4$, and with $\mathrm{Vol}=8\pi^2a^4/3$ we obtain $\int\GB\sqrt{g}=64\pi^2$ and $\chi=2$.

\emph{(ii) $S^2(a)\times S^2(b)$.} In this case $R=2/a^2+2/b^2$, $R_{ab}R^{ab}=2/a^4+2/b^4$ and $R_{abcd}R^{abcd}=4/a^4+4/b^4$, so $\GB=8/(a^2b^2)$, which is the product of the two Gaussian curvatures up to normalization. With $\mathrm{Vol}=16\pi^2a^2b^2$ we obtain $\int\GB\sqrt{g}=128\pi^2$ and $\chi=4$, in agreement with $\chi(S^2)^2=4$ and with the two bifurcation surfaces of the Nariai geometry.

\emph{(0) JT disk.} With $\rho_0$ as in Eq.~\eqref{eq:rho0} and the substitution $E=k^2/2$, symbolic integration gives Eq.~\eqref{eq:Zdisk}, while $S=\log Z+\beta\langle E\rangle$ gives Eq.~\eqref{eq:SJT}, from which $\partial_{S_0}S=1$ and $\partial_\beta\partial_{S_0}S=0$ follow identically.

\emph{(iii) Euclidean Schwarzschild.} With $\dd s^2=f\,\dd\tau^2+f^{-1}\dd r^2+r^2\dd\Omega^2$, $f=1-2M/r$, and $\tau$ of period $\beta=8\pi M$, the Ricci tensor vanishes and $\GB=R_{abcd}R^{abcd}=48M^2/r^6$. Integrating from $r=2M$ to infinity gives $\int_M\GB\sqrt{g}\,\dd^4x = 64\pi^2$, and hence $\chi=2$, while on the horizon $\oint{}^{(2)}\!R\,\dd A=8\pi=4\pi\chi(\Sigma)$.

\bibliography{refs}

\end{document}